\documentclass[prl,floatfix,showpacs,twocolumn,preprintnumbers,amsmath,amssymb,superscriptaddress]{revtex4}
\usepackage[sort&compress]{natbib}
\usepackage{graphicx,float,datetime}
\usepackage[ansinew]{inputenc}
\usepackage{hyperref, wasysym}
\usepackage{epstopdf, subfigure}

\newcommand*\xbar[1]{%
  \hbox{%
    \vbox{%
      \hrule height 0.5pt 
      \kern0.5ex
      \hbox{%
        \kern-0.1em
        \ensuremath{#1}%
        \kern-0.1em
      }%
    }%
  }%
} 

\newcommand{\udt}[3]{#1^{#2}_{\phantom{#2}#3}}

\newcommand{\dut}[3]{#1_{#2}^{\phantom{#2}#3}}
\newcommand{\dudt}[4]{#1_{#2\phantom{#3}#4}^{\phantom{#2}#3}}

\newdateformat{mydate}{\THEDAY\hspace{3pt}\monthname[\THEMONTH] \THEYEAR}

\allowdisplaybreaks[1]

\begin{document}

\begin{center}
\title{Growth factor in $f(T,\mathcal{T})$ gravity}
\date{\mydate\today}
\author{Gabriel Farrugia\footnote{gabriel.farrugia.11@um.edu.mt}}
\affiliation{Department of Physics, University of Malta, Msida, MSD 2080, Malta}
\affiliation{Institute of Space Sciences and Astronomy, University of Malta, Msida, MSD 2080, Malta}
\author{Jackson Levi Said\footnote{jackson.said@um.edu.mt}}
\affiliation{Department of Physics, University of Malta, Msida, MSD 2080, Malta}
\affiliation{Institute of Space Sciences and Astronomy, University of Malta, Msida, MSD 2080, Malta}

\begin{abstract}
{
We investigate the growth factor for sub-horizon modes during late times in $f(T,\mathcal{T})$ gravity, where $T$ is the torsion scalar and $\mathcal{T}$ is the trace of the stress-energy tensor. This is achieved by obtaining the modified M\'{e}sz\'{a}ros equation, which describes the evolution of the perturbations of the matter energy density, and obtaining numerical results. Such results are obtained by solving the modified continuity equation and analysing the behaviour of the solutions of the latter using various constraints on the integration constants. Furthermore, the role of the anisotropic term $\pi^{S}$ is investigated.
}
\end{abstract}


\maketitle

\end{center}

\section{I. Introduction}\label{sec:intro}

The theory of general relativity (GR) has been studied for over 100 years, but yet it fails to fully explain what the cause of the accelerated expansion of the universe is \cite{Riess:1998cb,Perlmutter:1998np,Hinshaw:2012aka,deSabbata:1990rn,Peebles:2002gy}. One of the prominent models used is the dark energy model which attempts to explain the late-time acceleration as a result of a kind of energy related to the cosmological constant. Alternative models and theories have tried to explain this phenomenon in the context of curvature, such as $f(R)$ gravity, modified Gauss-Bonnet gravity $f(G)$ and with a general coupling between the Ricci scalar and the Gauss-Bonnet terms in $f(R,G)$ gravity (see an extensive review on the cosmological implications in Refs. \cite{Nojiri:2010wj,DeFelice:2010aj} and references therein), and also via a coupling between matter and curvature through $f(R,T)$ gravity, where $T$ is the trace of the stress-energy tensor \cite{Harko:2011kv,Chakraborty:2012kj,Sun:2015yga,Singh:2016kek,Fayaz:2016bly}, among other approaches \cite{Clifton:2011jh}.

Recently, there has been an increase of interest in a different type of alternative theory of gravity, one which does not use curvature but uses torsion instead, which is called teleparallel gravity \cite{DeAndrade:2000sf,aldrovandi2012teleparallel}. This makes use of a different connection to the Christoffel symbol, called the Weitzenb\"{o}ck connection, which is a curvature-free quantity (which contrasts the Levi-Civita connection, used in curvature based models, which is a torsion-free quantity) and vierbein fields instead of a metric field. From this, torsion based quantities can be derived, most notably the torsion scalar $T$ (not to be confused with the trace of the stress-energy tensor), which can be used to explain gravity in terms of torsion. Thus, this torsion scalar replaces the idea of the Ricci scalar $R$. 

It was shown that the torsion formalism is equivalent to that of GR, called Teleparallel Equivalent General Relativity (TEGR), up to a boundary term difference  \cite{Garecki:2010jj,Maluf:2013gaa,Arcos:2005ec}. However, there are some subtle differences, such as having force-field equations analogues to those from electromagnetism instead of a geodesic equation, which allows the weak equivalence principle (WEP) to be violated, something which in GR is not possible \cite{Aldrovandi:2003xu,Aldrovandi:2003pa}. 

As GR was generalised to $f(R)$ gravity, teleparallel gravity was generalised to $f(T)$ gravity. Some interesting implications were discovered, such as allowing violations in the local Lorentz transformations (see \cite{Cai:2015emx} and references therein). However, recently it has been shown that local Lorentz invariance can be restored making the theory to be covariant as those in curvature models \cite{Krssak:2015lba}. Furthermore, in contrast to $f(R)$ gravity, the resulting field equations remain second order (in contrast to fourth order theories), making the field equations simpler to work with. Thus, although GR and TEGR are equivalent ways in describing gravity, $f(R)$ and $f(T)$ gravity models are fundamentally different, resulting in a new way to investigate the cosmological implications of such a theory \cite{Saez-Gomez:2016wxb}. Nonetheless, various investigations in $f(T)$ gravity have been applied within the realm of cosmology, including thermodynamics Ref. \cite{Salako:2013gka}, reconstruction Ref. \cite{Bamba:2012vg}, cosmological solutions Ref. \cite{Paliathanasis:2016vsw} and late-time acceleration Refs. \cite{Bengochea:2008gz,Linder:2010py} (detailed discussions on the topic of late-time acceleration can be found in Ref. \cite{Nesseris:2013jea}; for a detailed review on $f(T)$ gravity, see Ref. \cite{Cai:2015emx} and references therein).

In a similar way to what was done to $f(R,T)$ gravity (where here $T$ is the trace of the stress-energy tensor), $f(T)$ gravity can then be generalised into $f(T,\mathcal{T})$ gravity, where $\mathcal{T}$ is the trace of the stress energy tensor. The study of cosmological solutions in this theory has been investigated in Ref. \cite{Harko:2014aja}, as well as the aspect of reconstruction, thermodynamics and stability in Ref. \cite{Junior:2015bva}. It is also possible to investigate the introduction of the trace of the stress-energy tensor for non-linear couplings in curvature based gravity, say $f(G,\mathcal{T})$ gravity, which has been recently proposed in Ref. \cite{Sharif:2016xjv}. However, given the simplicity of the resulting field equations in the torsional perspective (of that being second order), it is much more reasonable to investigate the coupling between gravitation and matter through torsion rather than curvature (since the resulting field equations would be fourth order). This torsion and matter coupling further opens possibilities to describe what the nature of dark energy is, or more precisely what is causing the observed acceleration.

In this paper, we investigate the implications of such a theory in the realm of the growth evolution of the inhomogeneities of the universe during late times. This was originally considered by M\'{e}sz\'{a}ros for GR in 1974 \cite{Meszaros:1974tb}, and the perturbed equation is sometimes referred to as M\'{e}sz\'{a}ros equation or M\'{e}sz\'{a}ros effect \cite{dodelson2003modern}. This effect was investigated in various alternative and modified theories, including $f(R)$ Refs. \cite{Fu:2010zza,Bamba:2012qi,deMartino:2015zsa}, $f(T)$ gravity Refs. \cite{Zheng:2010am,Basilakos:2016xob}, and in holographic $f(T)$ gravity \cite{Karami:2011np}, yielding possible physical results when compared with $\Lambda$CDM models. Given the recent volume of work in the torsional description of gravity, we are interested in finding out what happens to the M\'{e}sz\'{a}ros effect in the $f(T,\mathcal{T})$ theory of gravity. In the case where $\mathcal{T}$ is introduced in $f(R,T)$ gravity, this was investigated by Alvarenga \textit{et. al} \cite{Alvarenga:2013syu}, and the growth factor was found to be dependent on the sub-horizon mode, which contrasts with what one obtains in GR and observational data. This puts in question the possibility of having such a theory as a possible candidate to explain late-time acceleration. Thus, we investigate whether such a result is also observed for $f(T,\mathcal{T})$ gravity. 

The paper is divided as follows. A brief overview of teleparallelism and $f(T,\mathcal{T})$ gravity is given in Sec. II, followed by the derivation of the modified M\'{e}sz\'{a}ros equation for the inhomogeneous universes in Sec. III. Afterwards, some solutions and also potential ansatz functions of the continuity equation are discussed in Sec. IV. Using these solutions, in Sec. V, numerical results for the growth factor are analysed. Finally, a conclusion about the results is given in Sec. VI.

\section{II. An overview of $f(T,\mathcal{T})$ gravity}\label{sec:field-equations}

\subsection{A. Connections, action and field equations}

In order to obtain a torsion based theory, one requires a new connection, the Weitzenb\"{o}ck connection $\widehat{\Gamma}^{\alpha}_{\mu\nu}$, which is defined as
\begin{equation}\label{eq:weitzenbockdef}
\widehat{\Gamma}^{\mu}_{\rho\nu}\equiv \dut{e}{a}{\rho}\partial_\mu \udt{e}{a}{\nu} + \dut{e}{a}{\rho}\udt{\omega}{a}{b\mu}\udt{e}{b}{\mu},
\end{equation}
where ${e^a}_\rho$ and ${e_a}^\mu$ are referred to as vierbeins (or tetrads) along with their respective inverses, and $\udt{\omega}{a}{b\mu}$ is called the \textit{purely inertial spin connection} which is related to the inertial effects of the system under consideration \cite{aldrovandi2012teleparallel,Krssak:2015lba}. Here, the Latin indices transform like a flat space coordinate, while the Greek indices transform like global coordinates. In this way, these vierbeins can be used to relate to the metric tensor $g_{\mu\nu}$ depending on the local position $x$ on the spacetime manifold by
\begin{equation}
g_{\mu\nu}\left(x\right)\equiv \udt{e}{a}{\mu}\left(x\right)\udt{e}{b}{\nu}\left(x\right)\eta_{ab},
\end{equation}
where $\eta_{ab}$ is the Minkowski metric tensor diag$(1,-1,-1,-1)$. In this way, the vierbein offer a way to connect the local Minkowski metric to the global metric tensor. From this point onward, the explicit expression of a local position $x$ will be suppressed. Using this connection, the \textit{torsion tensor} can be defined as
\begin{equation}\label{eq:torsiontensordef}
\udt{T}{\mu}{\rho\nu}\equiv \widehat{\Gamma}^{\mu}_{\nu\rho}-\widehat{\Gamma}^{\mu}_{\rho\nu}.
\end{equation}
The difference between the Weitzenb\"{o}ck and Levi-Civita connections is expressed by the contorsion tensor $\udt{K}{\alpha}{\mu\nu}$
\begin{equation}
\udt{K}{\alpha}{\mu\nu} \equiv \widehat{\Gamma}^{\alpha}_{\mu\nu}-\Gamma^{\alpha}_{\mu\nu},
\end{equation}
where $\Gamma^{\alpha}_{\mu\nu}$ is the Levi-Civita connection. The contorsion tensor can also be expressed in terms of the torsion tensor as
\begin{equation}\label{eq:contorsiondef}
\udt{K}{\lambda}{\mu\nu} = \frac{1}{2}\left(\dudt{T}{\mu}{\lambda}{\nu}+\dudt{T}{\nu}{\lambda}{\mu}-\udt{T}{\lambda}{\mu\nu}\right).
\end{equation}
In this way, the superpotential tensor can be defined as
\begin{equation}\label{eq:superpotentialdef}
\dut{S}{\rho}{\mu\nu}\equiv \frac{1}{2}\left(\udt{K}{\mu\nu}{\rho}+\delta_{\rho}^{\mu}\udt{T}{\alpha\nu}{\alpha}-\delta_{\rho}^{\nu}\udt{T}{\alpha\mu}{\alpha}\right).
\end{equation}
Using Eq. \eqref{eq:torsiontensordef} and \eqref{eq:superpotentialdef} leads to the torsion scalar
\begin{align}
T &\equiv \dut{S}{\rho}{\mu\nu}\udt{T}{\rho}{\mu\nu} \nonumber \\
& =\frac{1}{4}T^{\rho\mu\nu}T_{\rho\mu\nu}+\frac{1}{2}T^{\rho\mu\nu}T_{\nu\mu\rho}-\dut{T}{\rho\mu}{\rho}\udt{T}{\nu\mu}{\nu}, \label{eq:torsionscalardef}
\end{align}
which defines the action for teleparallel gravity to be
\begin{equation}\label{eq:teleparallel-action}
S = \dfrac{1}{16\pi G} \int d^4x \: e \: T + \int d^4x \: e \: \mathcal{L}_m,
\end{equation}
where $e = \det\left(\dut{e}{\mu}{A}\right) = \sqrt{-g}$ and $\mathcal{L}_m$ is the matter Lagrangian. As is done in GR, the torsion scalar in the action can be generalised to become a general function of both the torsion scalar and the trace of the energy momentum tensor $\mathcal{T}$, which results in
\begin{equation} \label{eq:general-action}
S = \dfrac{1}{16\pi G} \int d^4x \: e \: \left[T + f(T,\mathcal{T})\right] + \int d^4x \: e \: \mathcal{L}_m,
\end{equation}
where $f(T,\mathcal{T})$ represents the generalised function. This is the analogue of $f(R,T)$ gravity where $T$ is the trace of energy of the momentum tensor. By varying the action with respect to the inverse vierbein field $\delta \udt{e}{A}{\rho}$ (analogous to taking variations with the inverse metric tensor $\delta g^{\mu\nu}$ in the metric formalism in $f(R)$ gravity Ref. \cite{Sotiriou:2010mv}), the following field equations field equations are obtained
\begin{widetext}

\begin{align}
&\left(1+f_T\right)\left[e^{-1} \partial_\sigma \left(e \dut{S}{a}{\rho\sigma}\right)-\udt{T}{b}{\nu a}\dut{S}{b}{\nu\rho} + \udt{\omega}{b}{a\nu}\udt{S}{b}{\nu\rho}\right]
+ \left(f_{TT} \partial_\sigma T + f_{T\mathcal{T}} \partial_\sigma \mathcal{T}\right)\dut{S}{a}{\rho\sigma}
+ \dut{e}{a}{\rho} \left(\dfrac{T + f}{4}\right) \nonumber \\
&+ \dfrac{f_\mathcal{T}}{2} \left(\stackrel{\textbf{em}}{\dut{T}{a}{\rho}} + p\dut{e}{a}{\rho}\right)
= 4\pi G \stackrel{\textbf{em}}{\dut{T}{a}{\rho}}. \label{eq:general-field-equations}
\end{align}

\end{widetext}
where $\stackrel{\textbf{em}}{\dut{T}{\alpha}{\rho}}$ is the stress-energy tensor, which in terms of the matter Lagrangian is given by $\stackrel{\textbf{em}}{\dut{T}{\beta}{\rho}} = - e^{-1} \dfrac{\partial \left(e \mathcal{L}_m\right)}{\partial \udt{e}{a}{\rho}}$ (the full details of the derivation is given in Appendix I). In the case where the spin connection is zero, the field equations reduce to those found in Refs. \cite{Harko:2014aja,Saez-Gomez:2016wxb}.

It should be mentioned that the field equations listed in Refs. \cite{Harko:2014aja,Saez-Gomez:2016wxb} are derived within the pure vierbein formalism (the only dynamic variable is the vierbein) where the purely inertial spin connection, $\udt{w}{c}{ab}$, is assumed to vanish in all frames. This formulation results in having a breaking of the local Lorentz symmetry for $f(T) \neq T$ \cite{Ferraro:2006jd,Li:2010cg,Sotiriou:2010mv} due to previous assumption. This led to formulations of what are called good and bad tetrads (see Ref. \cite{Tamanini:2012hg} for more details). Recently, Kr\v{s}\v{s}\'{a}k and Saridakis show that this local Lorentz invariance problem can be solved by allowing a non-zero purely inertial spin connection, i.e. by taking the covariant formulation of the theory \cite{Krssak:2015oua}. Nonetheless, one can choose vierbeins which make the purely inertial spin connection vanish and still allow for local Lorentz invariance in this theory (such vierbeins are called proper vierbeins), whilst reducing the field equations to the standard pure vierbein ones. The vierbein considered in this paper is such type of vierbein and hence the field equations reduce to those in Refs. \cite{Harko:2014aja,Saez-Gomez:2016wxb} and will still be local Lorentz invariant. However, if a non-proper vierbein is chosen, one first requires to find the spin connection before deriving the field equations, which due to covariance, will be identical.

\subsection{B. Cosmologies in $f(T,\mathcal{T})$ gravity}

One can analyse some basic properties of the field equations by considering a spatially flat Friedmann-Lemaitre-Robertson-Walker (FLRW) metric
\begin{equation}\label{eq:flat-FLRW}
ds^2 = dt^2 - a^2(t)\left(dx^2+dy^2+dz^2\right),
\end{equation}
where $a(t)$ is the scale factor in terms of cosmic time. For such a metric, a diagonal vierbein field of the form
\begin{equation}\label{eq:diag-tetrad}
\dut{e}{\mu}{A} = \text{diag}\left(1,a,a,a\right),
\end{equation}
is considered. In this case, $T = -6H^2$. Using the field equations in Eq. \eqref{eq:general-field-equations}, this gives rise to the two GR modified equations
\begin{align}
& \left(1+f_T\right) 3H^2 + \dfrac{f+T}{4} + \dfrac{f_\mathcal{T}}{2} \left(\rho + p\right) = 4 \pi G \rho, \label{eq:00-zero} \\ 
& \left(1+f_T\right) \left(3H^2+\dot{H}\right) + \dfrac{f+T}{4} - 12H^2\dot{H} f_{TT} \nonumber \\
&+ H\left(\dot{\rho} - 3\dot{p}\right) f_{T\mathcal{T}} = -4 \pi G p. \label{eq:trace-zero}
\end{align}
These equations can be rearranged into a more familiar form
\begin{align}
& H^2 = \dfrac{8 \pi G}{3} \rho - 2H^2f_T - \dfrac{f}{6}- \dfrac{f_\mathcal{T}}{3} \left(\rho + p\right),  \\ 
&\dot{H} = -4 \pi G \left(p + \rho\right) + \dfrac{f_\mathcal{T}}{2} \left(\rho + p\right) \nonumber \\
&- H\left(\dot{\rho} - 3\dot{p}\right) f_{T\mathcal{T}} + 12H^2\dot{H} f_{TT} - \dot{H}f_T. \label{eq:dotH-equation}
\end{align}
Thus, by analysing the equations, one can define an effective dark energy (DE) pressure $p_{DE}$ and energy density $\rho_{DE}$ as follows
\begin{align}
&\dfrac{8\pi G}{3}\rho_{DE} \equiv - 2H^2f_T - \dfrac{f}{6}- \dfrac{f_\mathcal{T}}{3} \left(\rho + p\right), \label{eq:def-DE-energy-density} \\
&-4\pi G p_{DE} \equiv 4\pi G \rho_{DE}+\dfrac{f_\mathcal{T}}{2} \left(\rho + p\right) \nonumber \\
&- H\left(\dot{\rho} - 3\dot{p}\right) f_{T\mathcal{T}} + 12H^2\dot{H} f_{TT} - \dot{H}f_T, \label{eq:def-DE-pressure}
\end{align}
which in turn can be used to define an effective equation of state parameter $w_{DE}$ to be
\resizebox{\linewidth}{!}{
  \begin{minipage}{\linewidth}
  \begin{align}
&w_{DE} \equiv \dfrac{p_{DE}}{\rho_{DE}} \nonumber \\ 
&= -1 - \dfrac{2}{3}\dfrac{\dfrac{f_\mathcal{T}}{2} \left(\rho + p\right) - H\left(\dot{\rho} - 3\dot{p}\right) f_{T\mathcal{T}} + 12H^2\dot{H} f_{TT} - \dot{H}f_T}{- 2H^2f_T - \dfrac{f}{6}- \dfrac{f_\mathcal{T}}{3} \left(\rho + p\right)}.
\end{align}
  \end{minipage}
}
Note that this expression only makes sense given the provision that the denominator is non-zero. In this way, the effective equations become more familiar to the GR counterpart
\begin{align}
&H^2 = \dfrac{8 \pi G}{3} \left(\rho+\rho_{DE}\right),  \\ 
&\dot{H} = -4 \pi G \left(p + \rho + p_{DE} + \rho_{DE}\right). 
\end{align}
Together, they give rise to a modified continuity equation
\begin{equation}\label{eq:continuity-flat}
\dot{\rho} + \dot{\rho}_{DE} = -3H\left(\rho+p+\rho_{DE}+p_{DE}\right).
\end{equation}
Assuming that an equation of state for the matter pressure and density with equation of state parameter $w$, i.e. $p = w\rho$, the continuity equation reduces to
\begin{equation}
\dot{\rho} + \dot{\rho}_{DE} = -3H\left[\rho\left(1+w\right)+\rho_{DE}\left(1+w_{DE}\right)\right].
\end{equation}

At this point, one finds a coupling relation between the matter energy density and the effective DE energy density. In other words, this implies that the stress-energy tensor is not divergence free \cite{Saez-Gomez:2016wxb}. This occurs due to the matter and torsion coupling in the gravitational Lagrangian. In fact, removing such coupling restores the divergence free property in $f(T)$ gravity \cite{Zheng:2010am}. Other theories also result in such lack of divergence free property, for example $f\left(R,\mathcal{T}\right)$ gravity Ref. \cite{Harko:2011kv} and $f\left(R,\mathcal{L}_m\right)$ theories \cite{Harko:2008qz,Wang:2012rw}, where the coupling of matter and curvature is the cause of such divergencelessness. Having this not divergence free means that the standard GR continuity equation does not hold, and the matter evolution is influenced by this coupling. This influences standard fluid evolutions (e.g. photons energy density would necessarily evolve as $a^{-4}$), which are well defined from Maxwell-Boltzmann statistics \cite{dodelson2003modern}. Thus, this seemingly results into contradictions. However, this shortcoming can be resolved by choosing the right $f$ function which results into the stress-energy tensor to be truly divergenceless . This concept, formulated and used in Ref. \cite{Saez-Gomez:2016wxb}, allows to determine some of such possible functions. In this way, the Lagrangian is restricted. However, the full details of this approach are given in Section IV, where such solutions are extracted. In the subsequent sections however, this condition is not assumed to allow for generality.

For the time being, let us consider the particular case in which this effective dark energy fluid becomes an effective cosmological constant, i.e. one which requires the condition that $w_{DE} = -1$. In other words, the following condition must be satisfied
\begin{align}
0 &= \dfrac{f_\mathcal{T}}{2} \left(\rho + p\right) - H\left(\dot{\rho} - 3\dot{p}\right) f_{T\mathcal{T}} \nonumber \\
& + 12H^2\dot{H} f_{TT} - \dot{H}f_T. \label{eq:cosmological-constant-condition}
\end{align}
By rearranging Eq. \eqref{eq:dotH-equation}, the following expression is obtained
\begin{align}
&\dot{H}\left(1 - 12H^2 f_{TT} +f_T\right) = -4 \pi G \left(p + \rho\right) \nonumber \\
&+ \dfrac{f_\mathcal{T}}{2} \left(\rho + p\right) - H\left(\dot{\rho} - 3\dot{p}\right) f_{T\mathcal{T}},
\end{align}
which when combined with Eq. \eqref{eq:cosmological-constant-condition}, the following relation is obtained
\begin{align}
&-\dfrac{f_\mathcal{T}}{2} \rho \left(1 + w\right) + H\dot{\rho}\left(1 - 3w\right) f_{T\mathcal{T}} \nonumber \\
&= 4 \pi G \rho\left(1 + w\right)\left(2T f_{TT} + f_T\right),
\end{align}
where the equation of state and the torsion scalar $T = -6H^2$ have been used. One can note the dependence on $\dot{\rho}$ and $H$ in the equations, the former being dependent on the continuity equation whilst the latter can be expressed in terms of $T$. Since $\dot{\rho}$ depends also on $\dot{\rho}_{DE}$ and $\rho_{DE}$, one has to use the definition of this effective dark energy and combine with Eq. \eqref{eq:dotH-equation} to create a differential equation in terms of $T$ and $\mathcal{T}$ only [the $\rho$ terms can be expressed in terms of $\mathcal{T}$ by $\mathcal{T} = \rho(1-3w)$, except for the case $w = 1/3$]. This serves as a way to form a relation which gives rise to possible solutions of $f$ which effectively have the same effect as a cosmological constant. However, one can note one clear solution which is when $f$ is a constant, which results in the standard $\Lambda$CDM model. 

\section{III. Inhomogeneous evolution}

\subsection{A. Metric and Field Equations}

In order to analyse the evolution of the inhomogeneities of the universe, we shall consider the standard scalar perturbed FLRW metric (up to first order), which is of the form \cite{Pereira:2015pxa,ellis2012relativistic}
\begin{equation}
ds^2 = \left(1+2\phi\right)dt^2 - a^2(t)\left(1-2\psi\right)\delta_{ij} dx^i dx^j,
\end{equation}
for some scalar functions $\phi$ and $\psi$. Since the metric is generated by a vierbein field, one has to choose such a field which generates the above metric. One trivial choice would be
\begin{equation}
\dut{e}{\mu}{A} = \begin{pmatrix}
1+\phi & 0 \\
0 & a(1-\psi)\dut{\delta}{i}{m}
\end{pmatrix}.
\end{equation}
As it is argued in Zheng and Huang's paper Ref. \cite{Zheng:2010am}, in the case of $f(T)$ gravity, this results in compatibility issues with the integrated Sachs-Wolfe effect. For this reason, they proposed a non-diagonal vierbein field of the form
\begin{equation}
\dut{e}{\mu}{A} = \left(\delta^A_B + \dut{\chi}{B}{A}\right)\dut{\bar{e}}{\mu}{B},
\end{equation}
where $\dut{\bar{e}}{0}{A} = \delta^A_0$, $\dut{\bar{e}}{i}{A} = a \delta^A_i$ and 
\begin{equation}
\chi_{AB} = \begin{pmatrix}
\phi & \partial_i w \\
\partial_i \bar{w} & \delta_{ij} \psi + \partial_i \partial_j h + \epsilon_{ijk} \partial^k \tilde{h}
\end{pmatrix}.
\end{equation}
where $w$ and $\tilde{w}$ are two degrees of freedom of mass dimension and $h$ and $\tilde{h}$ are parity-violating terms. Using this vierbein, the following metric tensor is obtained
\begin{center}
\resizebox{0.95\linewidth}{!}{
  \begin{minipage}{\linewidth}
\begin{equation}
g_{\mu\nu} = \begin{pmatrix}
1+2\phi & a \partial_i \left(w + \tilde{w}\right) \\
a\partial_i \left(w + \tilde{w}\right) & -a^2\left[\left(1-2\psi\right)\delta_{ij} - 2\partial_i \partial_j h\right]
\end{pmatrix}.
\end{equation}
  \end{minipage}
}
\end{center}

To obtain the FLRW metric, a Newtonian gauge is considered, being $\tilde{w} = -w$ and $h = 0$. Having the vierbein field set, one can obtain the field equations from Eq. \eqref{eq:general-field-equations}. For more details about the quantities being considered, see Appendix II, where the veirbein field, superpotential and torsion tensors, torsion scalar and stress-energy tensor are all defined. It should be mentioned that in this paper the effect of the anisotropic term $\pi^{S}$ is considered. The equations are as follows:

\paragraph{(a)} Zero order equations: These are the ones found in Eqs. \eqref{eq:00-zero} and \eqref{eq:trace-zero}.
\paragraph{(b)} First order equations: In the following, the ``tensor" $\udt{E}{\rho}{A}$ corresponds to the free indices of the field equations, and hence the equation being considered are given by

\begin{widetext}
\begin{align}
\udt{E}{0}{0}: & \left(1+f_T\right)\left[a^{-2} \partial^2 \psi - 3H \dot{\psi} - 3H^2\phi\right] + 3H^2 \bigg[f_{TT}\left(12H\dot{\psi} + 12H^2\phi - 4a^{-1} H \partial^2 w\right) \nonumber \\
&+ f_{T\mathcal{T}}\left(\delta\rho - 3\delta p - \partial^2 \pi^{S}\right)\bigg] + \dfrac{f_{\mathcal{T}}}{4} \left(\delta\rho - 3\delta p - \partial^2 \pi^{S}\right) + f_{\mathcal{T}} \dfrac{\delta \rho + \delta p}{2} + \dfrac{\rho + p}{2}\bigg[f_{T\mathcal{T}}\Big(12H\dot{\psi} \nonumber \\ 
&+12H^2 \phi - 4a^{-1}H\partial^2 w\Big) + f_{\mathcal{T}\mathcal{T}}\left(\delta\rho - 3\delta p - \partial^2 \pi^{S} \right) \bigg] = 4\pi G \delta\rho \label{eq:00-first} \\
\udt{E}{i}{0}: & - a^{-2} \partial^i \left(\dot{\psi} + H\phi\right)\left(1+f_T\right) -a^{-2} \partial^i \psi  \left(-12H\dot{H} f_{TT} + \left(\dot{\rho} - 3\dot{p}\right) f_{T\mathcal{T}} \right)+ \dfrac{f_\mathcal{T}}{2} \left(\rho + p\right) \partial^i v \nonumber \\
& = 4 \pi G \left(\rho + p\right) \partial^i v, \label{eq:i0-first} \\
\udt{E}{0}{i}: & \left(1+f_T\right)\left[a^{-1}H\partial_i \phi + a^{-1}\partial_i \dot{\psi}\right] - a^{-1} H \bigg[f_{TT}\left(12H\partial_i \left(\dot{\psi}+H\phi\right)-4a^{-1}H\partial_i \partial^2 w\right) \nonumber \\
&+ f_{T\mathcal{T}}\partial_i\left(\delta\rho - 3\delta p - \partial^2 \pi^{S}\right)\bigg]-\dfrac{a}{2}f_\mathcal{T} \left(\rho+p\right) \partial_i v = -4\pi G a \left(\rho+p\right)\partial_i v, \label{eq:0i-first} \\
Tr\left(\udt{E}{i}{j}\right): & \left(1+f_T\right)\left[H\dot{\phi}+3H^2\phi+3H\dot{\psi}+2\dot{H}\phi+\ddot{\psi}-\dfrac{1}{3}a^{-2}\partial^2\left(\psi-\phi\right)\right]+f_{TT}\Big(-36H^3\dot{\psi}-36H^4\phi \nonumber \\
&-36H\dot{H}\dot{\psi}-60H^2\dot{H}\phi-12H^2\ddot{\psi}-12H^3\dot{\phi}+8a^{-1}H^3\partial^2 w + 12H\dot{H}a^{-1}\partial^2 w +4a^{-1}H^2\partial^2\dot{w}\Big) \nonumber \\
& - \dfrac{f_{\mathcal{T}}}{4}\left(\delta\rho - 3\delta p - \dfrac{5}{3}\partial^2 \pi^{S}\right) + f_{T\mathcal{T}}\bigg[\left(-3H^2-\dot{H}\right)\left(\delta\rho - 3\delta p - \partial^2 \pi^{S}\right) - H\big(\delta\dot{\rho} - 3\delta\dot{p} \nonumber \\
&- \partial^2 \dot{\pi}^{S}\big) + \left(\dot{\rho}-3\dot{p}\right)\left(\dot{\psi}+2H\phi-\dfrac{1}{3}a^{-1}\partial^2 w\right)\bigg] + 12H^2\dot{H}\Big[12H\left(\dot{\psi}+H\phi\right) \nonumber \\
&-4a^{-1}H\partial^2 w\Big]f_{TTT} + f_{TT\mathcal{T}}\Big\lbrace 12H^2\dot{H}\left(\delta\rho - 3\delta p - \partial^2 \pi^{S}\right)-H\left(\dot{\rho}-3\dot{p}\right)\Big[12H\left(\dot{\psi}+H\phi\right) \nonumber \\
&-4a^{-1}H\partial^2 w\Big]\Big\rbrace -H\left(\dot{\rho}-3\dot{p}\right)\left(\delta\rho - 3\delta p - \partial^2 \pi^{S}\right)f_{T\mathcal{T}\mathcal{T}}  = 4\pi G\left(\delta p +\dfrac{\partial^2 \pi^{S}}{3}\right), \label{eq:trace-first} \\
\udt{E}{i}{j}, \: i \neq j: & \dfrac{1}{2} a^{-2} \left(1+f_T\right) \partial_j \partial^i \left(\phi-\psi\right)-\dfrac{1}{2} a^{-1}\partial_j \partial^i w \left(-12H\dot{H}f_{TT} + \left(\dot{\rho}-3\dot{p}\right)f_{T\mathcal{T}}\right) - \dfrac{1}{2}f_{\mathcal{T}} \partial_j\partial^i \pi^{S} \nonumber \\
&= -4\pi G\partial_j \partial^i \pi^{S}. \label{eq:ij-first}
\end{align}
\end{widetext}

Similar to what was obtained by Zheng and Huang, the Parity-violating term $\tilde{h}$ vanishes whilst the $w$ term survives \cite{Zheng:2010am}. Similar to Harko \textit{et. al}, the existence of the anisotropic term $\pi^{S}$ and the fractional energy density and pressure are retained in the equations \cite{Harko:2014aja}. These are always coupled with derivatives of $\mathcal{T}$ except for $\pi^{S}$ in the last equation which is independent of the function $f$. In this case, the GR equations are obtained for $f(T,\mathcal{T}) = 0$ and $\pi^{S} = 0$. 

\subsection{B. Conservation Equations - Continuity and Velocity}\label{sec:conservation-equations}

The following conservation equations were derived using the field equations. Nonetheless, these can be obtained by taking the divergence of the stress-energy tensor. 

\paragraph{(a)} Continuity equation: The zero-order form of the continuity equation is the one given in the previous section, Eq. \eqref{eq:continuity-flat}. Substituting for the effective dark energy fluid yields
\begin{align}
&4\pi G \left[\dot{\rho}+3H\left(\rho+p\right)\right] = \dfrac{3H}{2}f_{\mathcal{T}}\left(\rho+p\right) + \dfrac{f_\mathcal{T}}{4} \big(3\dot{\rho} \nonumber \\
&-\dot{p}\big) - 6H\dot{H}\left(\rho+p\right)f_{T\mathcal{T}} + \dfrac{\rho+p}{2}\left(\dot{\rho}-3\dot{p}\right)f_{\mathcal{T}\mathcal{T}}. \label{eq:continuity-zero}
\end{align}
On the other hand, the first order equation is

\begin{widetext}

\begin{align}
&4\pi G \left[\delta\dot{\rho} + 3H\left(\delta\rho + \delta p + \dfrac{\partial^2 \pi^{S}}{3}\right) - 3\left(\rho + p\right)\dot{\psi} + \left(\rho + p\right)\partial^2 v\right] = f_{\mathcal{T}} \bigg[\dfrac{\rho + p}{2} \left(\partial^2 v - 3 \dot{\psi}\right) + \dfrac{H}{2} \big(3 \delta \rho \nonumber \\
&+ 3 \delta p + \partial^2 \pi^{S}\big) + \dfrac{1}{4} \left(3\delta\dot{\rho} - \delta\dot{p} - \partial^2 \dot{\pi^{S}}\right)\bigg] + f_{T\mathcal{T}} \bigg[-a^{-2}\partial^2 \psi \left(\dot{\rho}-3\dot{p}\right) + 2\left(\rho+p\right)\big(9H^2\dot{\psi}+9H^3 \phi \nonumber \\
& - 2a^{-1}H^2\partial^2 w + 3\dot{H}\dot{\psi} + 3H \ddot{\psi} + 6H\dot{H}\phi + 3H^2\dot{\phi} - a^{-1}H \partial^2 \dot{w} - a^{-1} \dot{H} \partial^2 w \big) - 6H\dot{H}\left(\delta\rho+\delta p\right) \nonumber \\
&+ 3H\left(H+\dot{\psi}\right) \left(\dot{\rho}-3\dot{p}\right) + \left(\dot{\rho}+\dot{p}\right)\left(6H\dot{\psi}+6H^2\phi-2a^{-1}H\partial^2 w\right) \bigg] + f_{\mathcal{T}\mathcal{T}} \bigg[\dfrac{3H}{2}\left(\rho + p\right)\big(\delta\rho - 3\delta p \nonumber \\
& -\partial^2 \pi^{S}\big) + \dfrac{1}{4}\left(3\delta\rho - \delta p -\partial^2 \pi^{S}\right)\left(\dot{\rho}-3\dot{p}\right) + \dfrac{1}{2}\left(\rho + p\right)\left(\delta\dot{\rho} - 3\delta \dot{p} -\partial^2 \dot{\pi^{S}}\right) + \dfrac{1}{2}\left(\dot{\rho}+\dot{p}\right)\big(\delta\rho - 3\delta p \nonumber \\
&-\partial^2 \pi^{S}\big)\bigg] -6H\dot{H} f_{TT\mathcal{T}} \left(\rho + p\right) \left(12H\dot{\psi} + 12H^2\phi -4a^{-1} H \partial^2 w\right) + \dfrac{f_{\mathcal{T}\mathcal{T}\mathcal{T}}}{2}\left(\rho + p\right)\left(\dot{\rho}-3\dot{p}\right)\big(\delta\rho - 3\delta p \nonumber \\
&-\partial^2 \pi^{S}\big) + \dfrac{f_{T\mathcal{T}\mathcal{T}}}{2}\left(\rho + p\right) \left[\left(\dot{\rho}-3\dot{p}\right)\left(12H\dot{\psi}+12H^2\phi-4a^{-1}H\partial^2 w\right) - 12H\dot{H}\left(\delta\rho - 3\delta p -\partial^2 \pi^{S}\right)\right] \label{eq:continuity-first}
\end{align}
\paragraph{(b)} Velocity equation: Since the velocity is a first order quantity, there is not a zeroth order equation, but instead only a first order one, which is given to be
\begin{align}
&-4\pi G \left[a^2\left(\rho+p\right)\partial_i \dot{v} + a^2 \dot{p}\partial_i v + 2a^2 H\left(\rho+p\right)\partial_i v + \left(\rho+p\right)\partial_i \phi + \partial_i \delta p + \partial_i \partial^2 \pi^{S}\right] = -\dfrac{1}{2}a^2 f_\mathcal{T} \big(\rho \nonumber \\
&+p\big)\partial_i \dot{v} + \dfrac{1}{4} a^2 f_\mathcal{T}\left(\dot{\rho}-3\dot{p}\right)\partial_i v - a^2 H f_\mathcal{T} \left(\rho+p\right) \partial_i v - \dfrac{1}{2}f_\mathcal{T} \left(\rho+p\right) \partial_i \phi + \dfrac{1}{4}f_\mathcal{T} \partial_i \left(\delta\rho - 3\delta p - 3\partial^2 \pi^{S}\right). \label{eq:velocity}
\end{align}
\end{widetext}

\subsection{C. Deriving the $f(T,\mathcal{T})$ M\'{e}sz\'{a}ros equation}\label{sec:overdensity-solution}

In what follows, we investigate how the inhomogeneous structure grows during matter dominated eras, i.e. our interest lies in what happens during the matter dominated era ($p = \delta p = 0$). This is achieved by investigating sub-horizon modes ($k >> aH$) and obtaining the $f(T,\mathcal{T})$ equivalent of the M\'{e}sz\'{a}ros equation. To make the calculations simpler, the equations will be Fourier transformed but no new symbols shall be applied to avoid confusion. 

\subsubsection{I. Sub-horizon approximations}

We start off by defining the gauge invariant fractional matter perturbation $\delta_m$ to be given by
\begin{equation}\label{eq:gauge-invariant-quantity}
\delta_m \equiv \dfrac{\delta \rho}{\rho} - 3Ha^2 v.
\end{equation}
From Eq. \eqref{eq:i0-first} and \eqref{eq:0i-first} 
\begin{align}
&f_{TT} \left(12H^2 \dot{\psi} + 12H^3 \phi - 12H\dot{H}\psi - 4a^{-1}H^2 k^2 w\right) \nonumber \\
&= f_{T\mathcal{T}} \left[-H\left(\delta \rho - 3\delta p -k^2\pi^{S}\right)-\psi\left(\dot{\rho}-3\dot{p}\right)\right].
\end{align}
To eliminate $\phi$ from the equation, we use Eq. \eqref{eq:ij-first}
\begin{align}
&\phi\left(1+f_T\right) = \psi\left(1+f_T\right) + aw\bigg[-12H\dot{H}f_{TT} + \big(\dot{\rho} \nonumber \\
&-3\dot{p}\big)f_{T\mathcal{T}}\bigg] + a^2 \pi^{S} \left(f_{\mathcal{T}}-8\pi G\right),
\end{align}
which when combined yields
\begin{widetext}
\begin{align}
&\left(1+f_T\right)f_{TT}\left(12H^2\dot{\psi}-12H\dot{H}\psi - 4a^{-1}H^2 k^2 w\right) +  12H^3f_{TT}\bigg\lbrace \psi\left(1+f_T\right)+aw\bigg[-12H\dot{H}f_{TT}+\big(\dot{\rho} \nonumber \\
&-3\dot{p}\big)f_{T\mathcal{T}}\bigg]+a^2 \pi^{S}\left(f_\mathcal{T} - 8\pi G\right) \bigg\rbrace =  \left(1+f_T\right)f_{T\mathcal{T}} \left[-H\left(\delta \rho - 3\delta p -k^2\pi^{S}\right)-\psi\left(\dot{\rho}-3\dot{p}\right)\right]. \label{eq:sub-horizon-1}
\end{align}
\end{widetext}

In the case of sub-horizon modes, the equation reduces to
\begin{equation}\label{eq:sub-horizon-2}
H^2 \psi + \delta \rho \sim k^2 \left(a^{-1} H w + \pi^{S}\right).
\end{equation}
However, from the definition of $\delta_m$ Eq. \eqref{eq:gauge-invariant-quantity},  one finds that the its order is
\begin{equation}
\delta_m \sim \dfrac{\delta \rho}{H^2} + H a^2 v.
\end{equation}
Thus, we get
\begin{equation}\label{eq:sub-horizon-3}
H^2 \psi + H^2 \delta_m + H^3 a^2 v \sim k^2 \left(a^{-1} H w + \pi^{S}\right).
\end{equation}
Note that this condition is only true if $f$ is also dependent on $\mathcal{T}$. For $f = f(T)$, the $\delta\rho$ term is not present in Eq. \eqref{eq:sub-horizon-1}, which results into
\begin{equation}
H^2 \psi \sim k^2 a^{-1} H w.
\end{equation}
However, since we are mostly interested in what happens when the function $f$ is also dependent on $\mathcal{T}$, this detail shall be ignored (the details for $f(T)$ gravity can be found in Zheng and Huang's paper \cite{Zheng:2010am}).

Using the first order trace equation Eq. \eqref{eq:trace-first}, its order is
\begin{equation}\label{eq:trace-order}
\phi \sim \psi + aHw + a^2 \pi^{S}.
\end{equation}
On the other hand, for a matter dominated universe, the velocity equation Eq. \eqref{eq:velocity} reduces to
\begin{align}
&\dot{v} + 2Hv + \dfrac{\phi}{a^2} + \dfrac{k^2 \pi^{S}}{a^2 \rho} = \dfrac{f_{\mathcal{T}}}{4\pi G a^2 \rho} \bigg[\dfrac{1}{2}a^2 \rho \dot{v} \nonumber \\
&- \dfrac{1}{4} a^2 \dot{\rho} v + \dfrac{1}{2} \rho \phi - \dfrac{1}{4} \rho \delta_m + \dfrac{1}{4}a^2 \rho v H + \dfrac{3}{4} k^2 \pi^{S}\bigg], \label{eq:velocity-matter}
\end{align}
whose order is given by
\begin{equation}
Hv + \dfrac{\phi}{a^2} + \dfrac{k^2 \pi^{S}}{a^2 H^2} \sim \dfrac{\delta_m}{a^2}.
\end{equation}
For the $f = f(T)$ case, the right hand side (RHS) of Eq. \eqref{eq:velocity-matter} is zero, and hence the order becomes
\begin{equation}
Hv + \dfrac{\phi}{a^2} \sim \dfrac{k^2 \pi^{S}}{a^2 H^2}.
\end{equation}
Now, combining with Eq. \eqref{eq:sub-horizon-2}
\begin{equation}
\dfrac{\phi}{a^2} + \dfrac{\psi}{a^2} \sim k^2 \left(\dfrac{w}{a^3 H} + \dfrac{\pi^{S}}{a^2 H^2}\right),
\end{equation}
which when combined with Eq. \eqref{eq:trace-order} yields
\begin{equation}
\dfrac{\psi}{a^2} \sim \dfrac{Hw}{a} + a \pi^{S} + k^2 \left(\dfrac{w}{a^3 H} + \dfrac{\pi^{S}}{a^2 H^2}\right).
\end{equation}
Therefore, in the sub-horizon limit, one concludes that
\begin{equation}
\dfrac{\psi}{a^2} >> \dfrac{Hw}{a} + \pi^{S}.
\end{equation}

By Eq. \eqref{eq:trace-order}, this implies $\phi \simeq \psi$. Note that from the previous relation, $\psi >> a^2 \pi^{S}$. Lastly, by Eq. \eqref{eq:i0-first} and taking its order, one concludes
\begin{equation}
H\phi + H\psi \sim a^2 H^2 v \implies Hv \sim \dfrac{\psi}{a^2} >> \dfrac{Hw}{a}+\pi^{S}.
\end{equation}

\subsubsection{II. The M\'{e}sz\'{a}ros equation}
Using Eq. \eqref{eq:00-first}, \eqref{eq:0i-first} and the definition of the $\delta_m$, one finds
\begin{align}
&4\pi G \delta_m = \left(1+f_T\right) \dfrac{k^2 \psi}{a^2 \rho} - \dfrac{3}{2} a^2 f_\mathcal{T} Hv \nonumber \\
&+ \dfrac{1}{2} \bigg\lbrace f_{T\mathcal{T}} \left[12H\left(\dot{\psi}+H\phi\right)-4a^{-1}Hk^2 w\right] \nonumber \\
&+ f_{\mathcal{T}\mathcal{T}}\left(\rho \delta_m + 3Ha^2 \rho v - k^2 \pi^{S}\right)\bigg\rbrace + \dfrac{f_{\mathcal{T}}}{4\rho}\big(3\rho\delta_m \nonumber \\
&- 9Ha^2\rho v - k^2 \pi^{S}\big). 
\end{align}
Using the subhorizon relationships, this reduces to
\begin{equation}\label{eq:delta-relation-1}
\left(4\pi G - \dfrac{3}{4}f_{\mathcal{T}} - \dfrac{1}{2}f_{\mathcal{T}\mathcal{T}} \rho\right)\delta_m = \left(1+f_T\right) \dfrac{k^2 \psi}{a^2 \rho}.
\end{equation}
For simplicity, we shall define the following quantity
\begin{equation}\label{eq:A-def}
A \equiv 4\pi G - \dfrac{3}{4}f_{\mathcal{T}} - \dfrac{1}{2}f_{\mathcal{T}\mathcal{T}} \rho.
\end{equation}
Now, from Eq. \eqref{eq:i0-first} and Eq. \eqref{eq:trace-first}, one finds 
\begin{align}
&4\pi G k^2 v - 12H^2\dot{H}f_{TT} \dfrac{k^2 w}{a\rho} = -\left(1+f_T\right)\dfrac{k^2 \dot{\psi}}{a^2 \rho} \nonumber \\
&+ 12H\dot{H}f_{TT} \dfrac{k^2 \psi}{a^2 \rho} - \dot{\rho} f_{T\mathcal{T}} \dfrac{k^2 \psi}{a^2\rho} + \dfrac{f_{\mathcal{T}}}{2}k^2 v + \big(8\pi G \nonumber \\
&- f_{\mathcal{T}}\big) k^2 \pi^{S} \dfrac{H}{\rho} - \left(1+f_T\right) \dfrac{k^2 \psi}{a^2 \rho} H - \dot{\rho}f_{T\mathcal{T}} \dfrac{k^2 w}{a \rho}H,
\end{align}
which for sub-horizon modes reduces to
\begin{align}\label{eq:velocity-reduced}
&4\pi G k^2 v = -\left(1+f_T\right)\dfrac{k^2 \dot{\psi}}{a^2 \rho} + 12H\dot{H}f_{TT} \dfrac{k^2 \psi}{a^2 \rho} \nonumber \\
&- \dot{\rho} f_{T\mathcal{T}} \dfrac{k^2 \psi}{a^2\rho} + \dfrac{f_{\mathcal{T}}}{2}k^2 v - \left(1+f_T\right) \dfrac{k^2 \psi}{a^2 \rho} H.
\end{align}

By differentiating with respect to time Eq. \eqref{eq:delta-relation-1}, and combining with the latter Eq. \eqref{eq:velocity-reduced} and using again Eq. \eqref{eq:delta-relation-1} gives
\begin{align}\label{eq:delta-relation-2}
&A \dot{\delta}_m + \left[\dot{A} + A\left(\dfrac{\dot{\rho}}{\rho}+3H\right)\right]\delta_m = \bigg(-4\pi G \nonumber \\
&+ \dfrac{f_\mathcal{T}}{2}\bigg)k^2 v.
\end{align}
Now, the sub-horizon limit of the velocity equation Eq. \eqref{eq:velocity-reduced} is
\begin{align}\label{eq:velocity-sub-horizon}
&\left(1-\dfrac{f_{\mathcal{T}}}{8\pi G}\right) \dot{v} + 2Hv + \left(1-\dfrac{f_{\mathcal{T}}}{8\pi G}\right)\dfrac{\phi}{a^2} + \bigg(1 \nonumber \\
&-\dfrac{3f_{\mathcal{T}}}{16\pi G}\bigg)\dfrac{k^2 \pi^{S}}{a^2\rho} = -\dfrac{f_{\mathcal{T}}}{16\pi G} \dfrac{\delta_m}{a^2}.
\end{align}
By differentiating with respect to time Eq. \eqref{eq:delta-relation-2}, combined with the previous equation Eq. \eqref{eq:velocity-sub-horizon} and the fact that $\phi \simeq \psi$ followed by Eq. \eqref{eq:delta-relation-1} and Eq. \eqref{eq:delta-relation-2} yields the modified M\'{e}sz\'{a}ros equation

\begin{widetext}
\begin{align}
&A \ddot{\delta}_m + \left\lbrace 2\dot{A} + A\left(\dfrac{\dot{\rho}}{\rho}+3H\right) + 2AH\left(1-\dfrac{f_{\mathcal{T}}}{8\pi G}\right)^{-1} + A \dfrac{d}{dt}\left[\ln\left(-4\pi G + \dfrac{f_{\mathcal{T}}}{2}\right)\right]\right\rbrace \dot{\delta}_m + \bigg\lbrace \dfrac{d}{dt}\bigg[\dot{A} \nonumber \\
&+A\left(\dfrac{\dot{\rho}}{\rho}+3H\right)\bigg] + 2H\left(1-\dfrac{f_{\mathcal{T}}}{8\pi G}\right)^{-1}\left[\dot{A}+A\left(\dfrac{\dot{\rho}}{\rho}+3H\right)\right] - 4\pi G \left(1-\dfrac{f_{\mathcal{T}}}{8\pi G}\right) \dfrac{\rho A}{1+f_T} - \dfrac{k^2 f_{\mathcal{T}}}{4a^2} \nonumber \\
&+ \dfrac{d}{dt}\left[\ln\left(-4\pi G + \dfrac{f_{\mathcal{T}}}{2}\right)\right]\left[\dot{A}+A\left(\dfrac{\dot{\rho}}{\rho}+3H\right)\right]\bigg\rbrace \delta_m - 4\pi G \left(1-\dfrac{3f_{\mathcal{T}}}{16\pi G}\right) \dfrac{k^4 \pi^{S}}{a^2 \rho} = 0. \label{eq:modified-Meszaros}
\end{align}
\end{widetext}

Let us analyse some of the properties of this evolution equation. One immediately notes that the equation is now dependent on the sub-horizon mode $k$, present in the two terms
\begin{align}
&\dfrac{k^2 f_{\mathcal{T}}}{4a^2}, &&4\pi G \left(1-\dfrac{3f_{\mathcal{T}}}{16\pi G}\right) \dfrac{k^4 \pi^{S}}{a^2 \rho}.
\end{align}
This contrasts from the GR M\'{e}sz\'{a}ros equation, but agrees with the $f(R,T)$ model \cite{Alvarenga:2013syu}. For this to make sense, we either require each term to vanish or their sum to vanish. In the first case, the following conditions need to be met
\begin{equation}
\pi^{S} = 0, \: f_\mathcal{T} = 0.
\end{equation}
Having $\pi^{S} = 0$ means that no fluid anisotropy exists while $f_\mathcal{T} = 0$ implies that $f = f(T)$, in other words, the theory reduces reduces to standard $f(T)$ gravity models.

In the second case, this leads to the following relation
\begin{equation}\label{eq:pis-relation}
\pi^{S} = -\dfrac{1}{k^2} \dfrac{\rho f_{\mathcal{T}} \delta_m}{16\pi G - 3 f_{\mathcal{T}}}.
\end{equation}
Recall that this $\pi^{S}$ is not the same anisotropy term defined in the stress-energy tensor, but its Fourier transform. This couples the effect of the trace of the stress-energy tensor with the anisotropic term. Furthermore, if $\pi^{S} = 0$, this leads to $f_\mathcal{T} = 0$, and hence to the standard $f(T)$ model as before. One also notes the dependence of $\pi^{S}$ to be inversely proportional to $k^2$, and hence decreases for larger values of $k$, in which for sub-horizon modes leads to the effect of anisotropy to be small. This makes sense since the effect of anisotropy should be small. Furthermore, one can also note the order of this expression to be
\begin{equation}
\pi^{S} \sim \dfrac{H^2 \delta_m}{k^2},
\end{equation}
which is in line with Eq. \eqref{eq:sub-horizon-3}. However, one should carefully interpret this result. The modified M\'{e}sz\'{a}ros equation was obtained under the approximation of sub-horizon modes, so this equality only holds for such approximations. If one solves the evolution exactly, the form of $\pi^{S}$ will most likely change, but it should reduce to Eq. \eqref{eq:pis-relation} for sub-horizon modes. Nonetheless, one can use this relation in two different ways. The first is by letting the parameters of the function $f$ set the value of $\pi^{S}$, while the second is by setting a form of $\pi^{S}$, which leads to a constraint on the parameters of $f$. These are investigated in Section V.

Lastly, in the case of $f(T)$ gravity, we have $A = 4\pi G$ and $\dot{\rho} + 3H\rho = 0$ (by continuity equation Eq. \eqref{eq:continuity-zero}). This reduces the equation to
\begin{equation}
\ddot{\delta}_m + 2H \dot{\delta}_m - \dfrac{4\pi G}{1+f_T}\rho \delta_m = \dfrac{k^4 \pi^{S}}{a^2 \rho}.
\end{equation}
This agrees with Zheng and Huang's result except for the $\pi^{S}$ term, which survives even for TEGR (i.e. $f = 0$) \cite{Zheng:2010am}. This shows that the evolution is dependent on the sub-horizon mode even in TEGR provided that the universe has non-zero anisotropy. Unless this term is inversely dependent on $k$ by at least $k^4$, this will lead to, eventually large, deviations for larger sub-horizon modes. 

\section{IV. Continuity equation solutions}\label{sec:continuity-solutions}

In this section, we investigate some possible solutions of the continuity equation Eq. \eqref{eq:continuity-zero}. Considering matter dominated times, the matter energy density should evolve as $\rho \propto a^{-3}$ (i.e. volumetric). For this reason, as discussed in Section II, the case in which the standard GR continuity equation is satisfied was considered. This simplifies the continuity equation into
\begin{equation}\label{eq:cont-diff-condition}
0 = f_{\mathcal{T}} + 8\dot{H}f_{T\mathcal{T}} + 2 \mathcal{T} f_{\mathcal{T}\mathcal{T}},
\end{equation}
where we have used the fact that for matter dominated universes, $\mathcal{T} = \rho$. Clearly, for any function $f(T,\mathcal{T}) = g(T)$, this is a solution of the differential equation. However, our main interest lies in the non-zero $\mathcal{T}$ solutions. Two such solutions have been found to satisfy the differential equation, which were also obtained by Diego \textit{et. al} \cite{Saez-Gomez:2016wxb}. 

\subsection{A. $f(T,\mathcal{T}) = g(\mathcal{T})$}

\noindent For such functions, Eq. \eqref{eq:cont-diff-condition} reduces to
\begin{equation}
0 = g^\prime + 2\mathcal{T} g^{\prime\prime},
\end{equation}
where primes are derivatives with respect to $\mathcal{T}$. This leads to the solution, $g(\mathcal{T}) = c_1 \sqrt{\mathcal{T}} + c_2$, where $c_1$ and $c_2$ are integration constants ($c_2$ represents a cosmological constant). The evolution of the universe for such a model can be seen by substituting in Eq. \eqref{eq:00-zero}, which gives
\begin{equation}\label{eq:00-zero-caseA.1}
H^2 = \dfrac{8\pi G}{3}\rho - \dfrac{c_1}{3}\rho^{1/2} - \dfrac{c_2}{6}.
\end{equation}
Since $c_1$ and $c_2$ are constants, by evaluating the equation for matter dominated times gives
\begin{equation}\label{eq:caseA-c1andc2-dependance}
c_1 = 3 \sqrt{\dfrac{8\pi G}{3{H_0}^2 \Omega_m}} \left[{H_0}^2\left(\Omega_m -1\right) - \dfrac{c_2}{6}\right],
\end{equation}
where the relations $\rho = \rho_0 a^{-3}$ with $\rho_0$ being a constant of proportionality, and the matter density parameter $\Omega_m$ being
\begin{equation}
\Omega_m \equiv \dfrac{\rho_0}{\rho_{c}} = \dfrac{8\pi G \rho_0}{3{H_0}^2},
\end{equation}
with $\rho_c$ being the critical density have been used. Substituting back into Eq. \eqref{eq:00-zero-caseA.1} yields
\begin{equation}\label{eq:00-zero-caseA.2}
H^2 = \frac{{H_0}^2 \Omega_m}{a^3}-\frac{c_2}{6}-\dfrac{1}{a^{3/2}} \left[{H_0}^2 (\Omega_m-1)-\frac{c_2}{6}\right].
\end{equation}
Thus, the evolution is only dependent on the parameter $c_2$. Note that $c_1 = 0$ gives the condition $c_2 = 6{H_0}^2 (\Omega_m-1)$, which is the standard $\Lambda$CDM model. With this in mind, one can define $c_2$ as
\begin{equation}\label{eq:caseA-epsilon}
c_2 \equiv 6{H_0}^2 (\Omega_m-1 + \epsilon),
\end{equation}
where $\epsilon$ is an arbitrary constant. This reduces Eq. \eqref{eq:00-zero-caseA.2} to 
\begin{equation}\label{eq:00-zero-caseA.3}
H^2 = {H_0}^2\left(\dfrac{\Omega_m}{a^3} - \Omega_m + 1 - \epsilon + \dfrac{\epsilon}{a^{3/2}}\right).
\end{equation}

\subsection{B. $f(T,\mathcal{T}) = Tg(\mathcal{T})$}

In this case, we consider a rescaling of the torsion scalar through some function of the trace of the stress-energy tensor. In this case, the Eq. \eqref{eq:cont-diff-condition} becomes
\begin{equation}
0 = g^\prime \left(1 - \dfrac{4\dot{H}}{3H^2}\right) + 2\mathcal{T}g^{\prime\prime}.
\end{equation}
From Eq. \eqref{eq:00-zero} and \eqref{eq:trace-zero}, one finds
\begin{equation}
\dfrac{\dot{H}}{H^2} = \dfrac{3}{1+g}\left(\mathcal{T}g^\prime - \dfrac{1+g}{2}\right),
\end{equation}
which when combined with the previous equation yields
\begin{equation}
0 = g^\prime \left(3-\dfrac{4\mathcal{T} g^\prime}{1+g}\right) + 2\mathcal{T}g^{\prime\prime}.
\end{equation}
This results in the following solution
\begin{equation}
g = -1-\dfrac{1}{\dfrac{2c_1}{\sqrt{\mathcal{T}}}+c_2},
\end{equation}
where $c_1$ and $c_2$ are integration constants. Note that in this case, the action for $f(T,\mathcal{T})$ gravity Eq. \eqref{eq:general-action} becomes
\begin{equation}
S = -\dfrac{1}{16\pi G} \int d^4x \: e \: \dfrac{T}{\dfrac{2c_1}{\sqrt{\mathcal{T}}}+c_2} + \int d^4x \: e \: \mathcal{L}_m.
\end{equation}
where the negative sign can be countered by assigning negative values to both the $c1$ and $c2$ parameters. For $c_1 = 0$, $c_2$ serves the role of a rescaling constant without a cosmological constant, with $c_2 = -1$ making it the standard teleparallel action. 

Now, by substituting this function in the energy density equation Eq. \eqref{eq:00-zero} yields
\begin{equation}
H^2 = -\dfrac{8\pi G}{3c_2} \left(2c_1 + c_2 \sqrt{\rho}\right)^2.
\end{equation}
Evaluating at today's time gives a way to define the constants $c_1$ and $c_2$ to the Hubble constant and today's energy density by
\begin{equation}\label{eq:00-zero-caseB.1}
{H_0}^2 = -\dfrac{8\pi G}{3c_2} \left(2c_1 + c_2 \sqrt{\rho_0}\right)^2.
\end{equation}
Hence, the expression for the Hubble parameter can be rewritten as
\begin{equation}\label{eq:00-zero-caseB.present-time}
\dfrac{H}{H_0} = \dfrac{2c_1 + c_2 \sqrt{\rho}}{2c_1 + c_2 \sqrt{\rho_0}}.
\end{equation}

Note that the constants relation can be rewritten as
\begin{equation}
{c_2}^2 \sqrt{\mathcal{T}_0} + c_2 \left(2c_1 \sqrt{\mathcal{T}_0} + \dfrac{3{H_0}^2}{8 \pi G}\right) + 4{c_1}^2 = 0,
\end{equation}
which is simply a quadratic in $c_2$, whose solutions are given by
\begin{equation}
c_2 = -\dfrac{c_1}{\sqrt{\mathcal{T}_0}} - \dfrac{1}{2 \Omega_m} \pm \dfrac{1}{2} \sqrt{\dfrac{4c_1}{\Omega_m \sqrt{\mathcal{T}_0}}+ \dfrac{1}{{\Omega_m}^2}}.
\end{equation}
This means that we do not have two degrees of freedom but essentially one, $c_1$. Note that although a quadratic in $c_2$ was arrived at, this can be treated vice-versa to be a quadratic in $c_1$ where $c_2$ becomes to degree of freedom. This solution for $c_2$ only produces for real roots, and hence we require
\begin{equation}
\dfrac{4c_1}{\Omega_m \sqrt{\mathcal{T}_0}}+ \dfrac{1}{{\Omega_m}^2} \geq 0 \implies c_1 \geq -\dfrac{\sqrt{\mathcal{T}_0}}{4 \Omega_m}.
\end{equation}
By defining $c_1$ to be
\begin{equation}\label{eq:caseB-epsilon}
c_1 \equiv -\dfrac{\epsilon \sqrt{\mathcal{T}_0}}{4\Omega_m},
\end{equation} 
for some constant $\epsilon$ which due to the condition for $c_1$ forces the condition $\epsilon \leq 1$, reduces the solution for $c_2$ to
\begin{equation}\label{eq:caseB-c1andc2-dependance}
c_2 = \dfrac{1}{2\Omega_m}\left(\dfrac{\epsilon}{2}-1 \pm \sqrt{1-\epsilon}\right).
\end{equation}
This in turn reduces Eq. \eqref{eq:00-zero-caseB.1} to
\begin{equation}\label{eq:00-zero-caseB.2}
\dfrac{H}{H_0} = \dfrac{-2\epsilon + a^{-3/2}\left(\epsilon-2\pm 2\sqrt{1-\epsilon}\right)}{-\epsilon-2\pm 2\sqrt{1-\epsilon}}.
\end{equation}
Thus, the evolution is only dependent on the magnitude of $\epsilon$. In the extremal case when $\epsilon = 1$, the evolution reduces to
\begin{equation}
\dfrac{H}{H_0} = \dfrac{2 + a^{-3/2}}{3}.
\end{equation}
For the case when $\epsilon = 0$ (i.e. $c_1 = 0$), $c_2$ has two solutions, $0$ or $-1/\Omega_m$. The first case becomes non-physical because Eq. \eqref{eq:00-zero-caseB.1} leads to $H = 0$. On the other hand, the second case reduces the equation to
\begin{equation}
\dfrac{H}{H_0} = a^{-3/2},
\end{equation}
which is precisely the evolution of a matter dominated universe. However, this only makes sense for $\Omega_m = 1$, since the action in this case is represented by
\begin{equation}
S = \int \Omega_m T + S_m.
\end{equation}
This is a rescaling scenario, and such rescaling should deviate slightly from TEGR (i.e. from $T$). Since only matter dominated universes are considered, this forces the matter density to be the only present component in the universe, and hence $\Omega_m = 1$ (this also follows for only matter universes in GR). Thus, since we are considering matter dominated universes, but closer to the observational value of $\Omega_m \approx 0.3$, such case would be non-physical. 

\subsection{C. Potential ansatz functions}

Other functions have been considered, however they prove to be inconsistent with the field equations. In particular, the following two have been considered.

\subsubsection{I. $f(T,\mathcal{T}) = \mathcal{T}g(T)$}

Since a rescaling of $T$ was considered, the converse is now assumed, i.e. a rescaling of $\mathcal{T}$. In this case, the continuity equation becomes
\begin{equation}
0 = g + 8 \dot{H}g^{\prime},
\end{equation}
where prime denotes a derivative with respect to $T$. Again, using Eq. \eqref{eq:00-zero} and \eqref{eq:trace-zero} gives the following relationship
\begin{equation}
\dot{H}\left(1+\mathcal{T}g^\prime + 2T\mathcal{T}g^{\prime\prime}\right) = \dfrac{T-\mathcal{T}g}{4},
\end{equation}
which when substituted in the previous equation yields
\begin{equation}\label{eq:cont-cond-1}
0 = g + g^\prime\left(2T-\mathcal{T}g\right)+2T\mathcal{T}gg^{\prime\prime}.
\end{equation}
To convert this equation into a differential equation of $T$ only, Eq. \eqref{eq:00-zero} was used to form a relation between $\mathcal{T}$ and $T$ which is
\begin{equation}
\mathcal{T}\left(4\pi G - \dfrac{3g}{4} + \dfrac{T g^\prime}{2}\right) = -\dfrac{T}{4}.
\end{equation}
The result of this is that Eq. \eqref{eq:cont-cond-1} is expressed as
\begin{align}
0 &= 16\pi G g - 3g^2 - 3Tgg^\prime + 32 \pi G T g^\prime \nonumber \\
&+ 4T^2 {g^\prime}^2 - 2T^2 gg^{\prime\prime}.
\end{align}

Solving this differential equation analytically is extremely difficult. However, one can note that $g$ being a constant is a possible solution. Assuming $g = c_1$, where $c_1$ is the constant, reduces the differential equation to
\begin{equation}
0 = 16\pi G c_1 - 3{c_1}^2.
\end{equation}
which gives two solutions, $c_1 = 0$ or $c_1 = 16\pi G/3$. In the first case, this means $g = 0 \implies f = 0$, which thus boils down to TEGR (without cosmological constant). In the second case, substituting in the energy density equation gives $T = 0 \implies H = 0$, which is not a physical solution. Since we are looking for a non-trivial solution, the only possible solution in this case would be the non-constant solution, which cannot be found analytically.  

However, one can analyse the differential equation using perturbation techniques by treating $T$ as a `first order' quantity. Assuming a solution of the form $g \approx g_0 + g_1 + g_2 + \dots$ leads to the following system of equations
\begin{align}
0 &= 16\pi G g_0- 3{g_0}^2, \\
0 &= 16\pi G g_1 - 3Tg_0{g_0}^{\prime} + 32\pi G T{g_0}^{\prime}, \\
0 &= 16\pi G g_2 - 3{g_1}^2 - 3T\left(g_0{g_1}^{\prime} + g_1{g_0}^{\prime}\right) \nonumber \\
&+ 32\pi G T{g_1}^{\prime} + 4T^2 {{g_0}^{\prime}}^2 - 2T^2 g_0 {g_0}^{\prime\prime} \\
&\hspace{3cm} \vdots \nonumber
\end{align} 
The first equation leads to $g_0 = 0$ or $g_0 = 16\pi G/3$ as before (these being the constant solutions). By substituting for the $g_1$ equation leads to $g_1 = 0$. Similarly, this leads to $g_2 = 0$, and so forth. Thus, the solution becomes $g \approx g_0$, which boils down to the two cases discussed previously. This might indicate that other solutions might not exist or such solution cannot be expanded as a power series solution, ultimately leading to no solutions for a possible rescaling of $\mathcal{T}$.

\subsubsection{II. $f(T,\mathcal{T}) = \alpha T^n \mathcal{T}^m$}

In this case, we consider the possibility of having a product solution, where $\alpha,n,m$ are constants. The continuity equation for this case reduces to
\begin{center}
\resizebox{0.92\linewidth}{!}{
  \begin{minipage}{\linewidth}
\begin{equation}
0 = \alpha m T^n \mathcal{T}^{m-1} \left(1 + 8\dot{H}n T^{-1} + 2\mathcal{T}(m-1) \mathcal{T}^{-1}\right).
\end{equation}
  \end{minipage}
}
\end{center}
This leads to the following possibilities, $\alpha = 0$, $m = 0$, $T = 0$, $\mathcal{T} = 0$ or the bracketed term to be zero. The first case reduces to TEGR (without cosmological constant), the second reduces to $f(T)$ gravity, while the third and the fourth give non-physical results. Thus, we consider the last, non-trivial case. This can be re-expressed as
\begin{equation}
2m-1-\dfrac{4\dot{H}}{3H^2} n = 0.
\end{equation}
However, since both $n,m$ are constants, this requires $\dot{H}/H^2$ to be constant. This simply leads to $H = H_0 a^{-3/2}$, which reduces the condition to
\begin{equation}\label{eq:n+m-condition-1}
2m+2n = 1.
\end{equation}

Note that this can only occur when $n \neq 0$ since when $n = 0$, no constraint on the evolution of $H$ would need to be set, and this sets $m = 1/2$. This simply reduces to the first solution encountered in Section IV. Thus, we shall now consider the case in which $n \neq 0$. In order for this condition to be consistent, the solution for $H$ has to be consistent with the field equations. Let us consider the energy density equation Eq. \eqref{eq:00-zero}, in which case, reduces to
\begin{equation}
H^2 = \dfrac{8\pi G}{3}\rho - \dfrac{\alpha T^n \mathcal{T}^m}{6}(2m-2n+1).
\end{equation}
Recall that $\rho = \rho_0 a^{-3}$. Since $\alpha$ is a constant independent of time, substituting for both $H$ and $\rho$ should make the equation independent of time. This leads to the following
\begin{align}
&{H_0}^2 a^{-3} = \dfrac{8\pi G}{3}\rho_0 a^{-3} \nonumber \\
&- \dfrac{\alpha}{6} (-6)^n {H_0}^{2n} {\rho_0}^{m} a^{-3n-3m}(2m-2n+1).
\end{align}
Since the scale factor is the only function which depends on time, all powers of $a$ must cancel. This sets another condition for $n,m$, being
\begin{equation}\label{eq:n+m-condition-2}
-3n -3m = -3 \implies n+m = 1.
\end{equation}
However, this contradicts the previous condition Eq. \eqref{eq:n+m-condition-1}. Thus, the only solutions are $\alpha \sqrt{\mathcal{T}}$ and the constant solution (obtained by taking both $n,m$ equal to 0).

\section{V. Numerical Results}\label{sec:numerical-results}

In this section, we shall consider various different scenarios concerning the evolution of the growth factor using the solutions found in the previous section. Since these solutions are based on the GR condition set on the continuity condition, the evolution differential equation Eq. \eqref{eq:modified-Meszaros} becomes

\begin{widetext}
\begin{align}
&A \ddot{\delta}_m + \left\lbrace 2\dot{A} + 2AH\left(1-\dfrac{f_{\mathcal{T}}}{8\pi G}\right)^{-1}+A \dfrac{d}{dt}\left[\ln\left(\dfrac{f_{\mathcal{T}}}{2}-4\pi G\right)\right]\right\rbrace \dot{\delta}_m + \bigg\lbrace \ddot{A} + 2\dot{A}H\left(1-\dfrac{f_{\mathcal{T}}}{8\pi G}\right)^{-1} \nonumber \\
&- 4\pi G \left(1-\dfrac{f_{\mathcal{T}}}{8\pi G}\right)^{-1} \dfrac{\rho A}{1+f_T} + \dot{A} \dfrac{d}{dt}\left[\ln\left(\dfrac{f_{\mathcal{T}}}{2}-4\pi G\right)\right] - \dfrac{k^2 f_{\mathcal{T}}}{4a^2} \bigg\rbrace \delta_m - 4\pi G \left(1-\dfrac{3f_{\mathcal{T}}}{16\pi G}\right) \dfrac{k^4 \pi^{S}}{a^2 \rho} = 0.
\end{align}
\end{widetext}

For simplicity, we define $D \equiv \delta_m(a)/\delta_m(a_i)$, where $a_i$ is some initial scale factor, in which it is considered to be 0.1. In the following, we analyse the evolution of $D$ with $a$. 

\subsection{A. Numerical Results for $\pi^{S} = 0$}

In the case where $\pi^{S} = 0$, the differential equation still remains dependent on $k$ due to the presence of the following term
\begin{equation}
\dfrac{k^2 f_{\mathcal{T}}}{4a^2} \delta_m.
\end{equation}
Thus, for any function considered here, unless $f$ is a function of torsion (or a cosmological constant), the evolution will be dependent on $k$. For this reason, one finds that for the functions considered in this paper, the evolution of $\delta_m$ will eventually cause either oscillations or accelerated growth as the value of $k$ changes, which is non-physical due to the fact that the evolution of $\delta_m$ should not change with the value of $k$ (or at least does not deviate from a $\Lambda$CDM solution by much for every sub-horizon $k$ value). 

\subsubsection{I. $f = c_1 \sqrt{\mathcal{T}} + c_2$}

From the previous section, it was found that $c_1$ and $c_2$ are dependent on each other through Eq. \eqref{eq:caseA-c1andc2-dependance}. Ultimately, the evolution Eq. \eqref{eq:00-zero-caseA.3} can be expressed by a single parameter $\epsilon$ defined in Eq. \eqref{eq:caseA-epsilon}. Thus, the evolution of $\delta_m$ can be analysed by varying the values of $\epsilon$.

\begin{figure}[h!]
\centering
\includegraphics[width=0.49\textwidth]{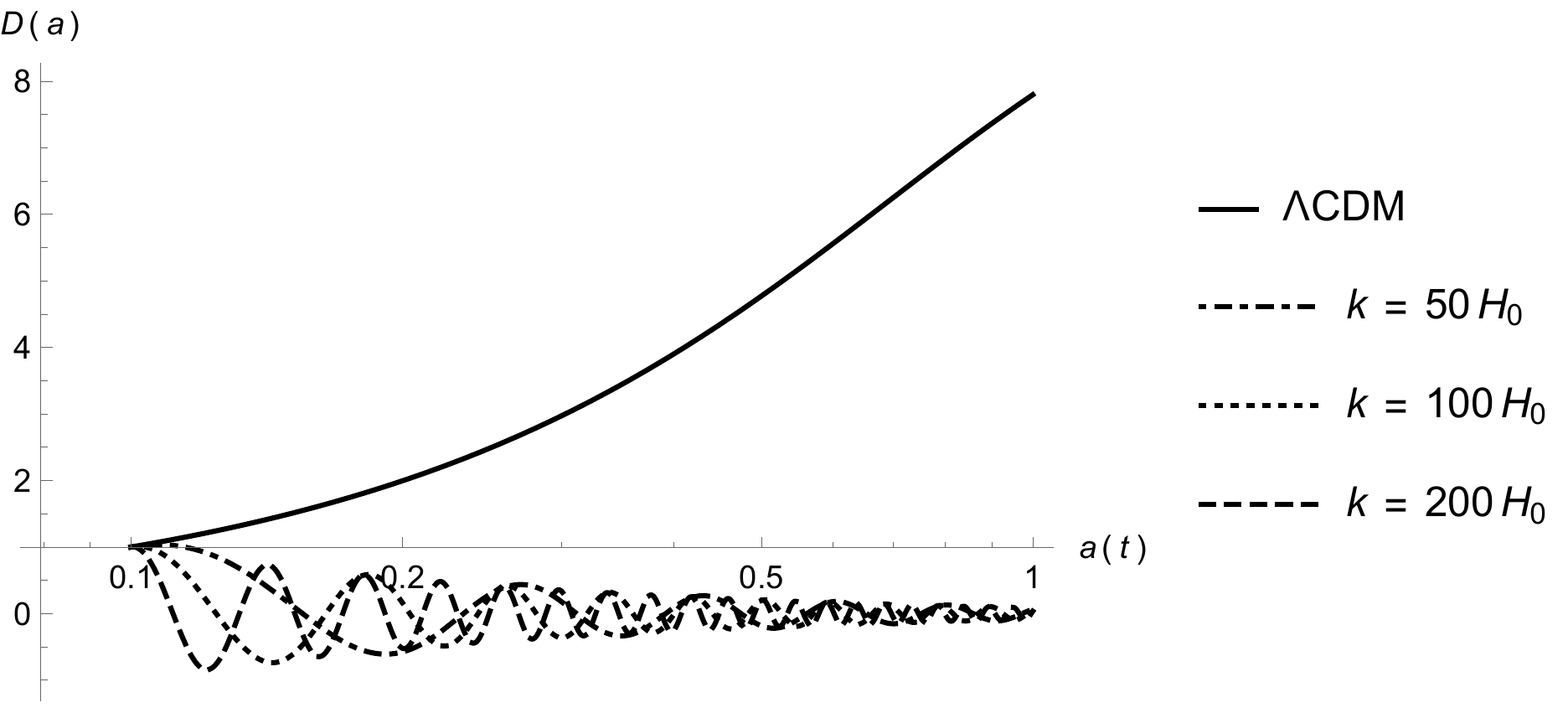}
\caption{Evolution for the $f = c_1 \sqrt{\mathcal{T}} + c_2$ model with $c_1$ and $c_2$ defined by Eqs. \eqref{eq:caseA-c1andc2-dependance} and \eqref{eq:caseA-epsilon} having $\epsilon = 1-\Omega_m$ (i.e. $c_2 = 0$). Note that the solution is very far from the $\Lambda$CDM solution and oscillatory. With increasing $k$, the number of periods increase.}
\label{fig1}
\end{figure}

The first case considered is where $\epsilon = 1-\Omega_m$ ($c_2 = 0$), which basically neglects the effect of the cosmological constant. As shown in Fig. \ref{fig1}, the effect of $k$ is already dominant, even for sufficiently small sub-horizon modes. Furthermore, the solution is oscillatory, with increasing periods for larger sub-horizon modes, and is far from the $\Lambda$CDM solution.

\begin{figure}[h!]
\centering
\includegraphics[width=0.49\textwidth]{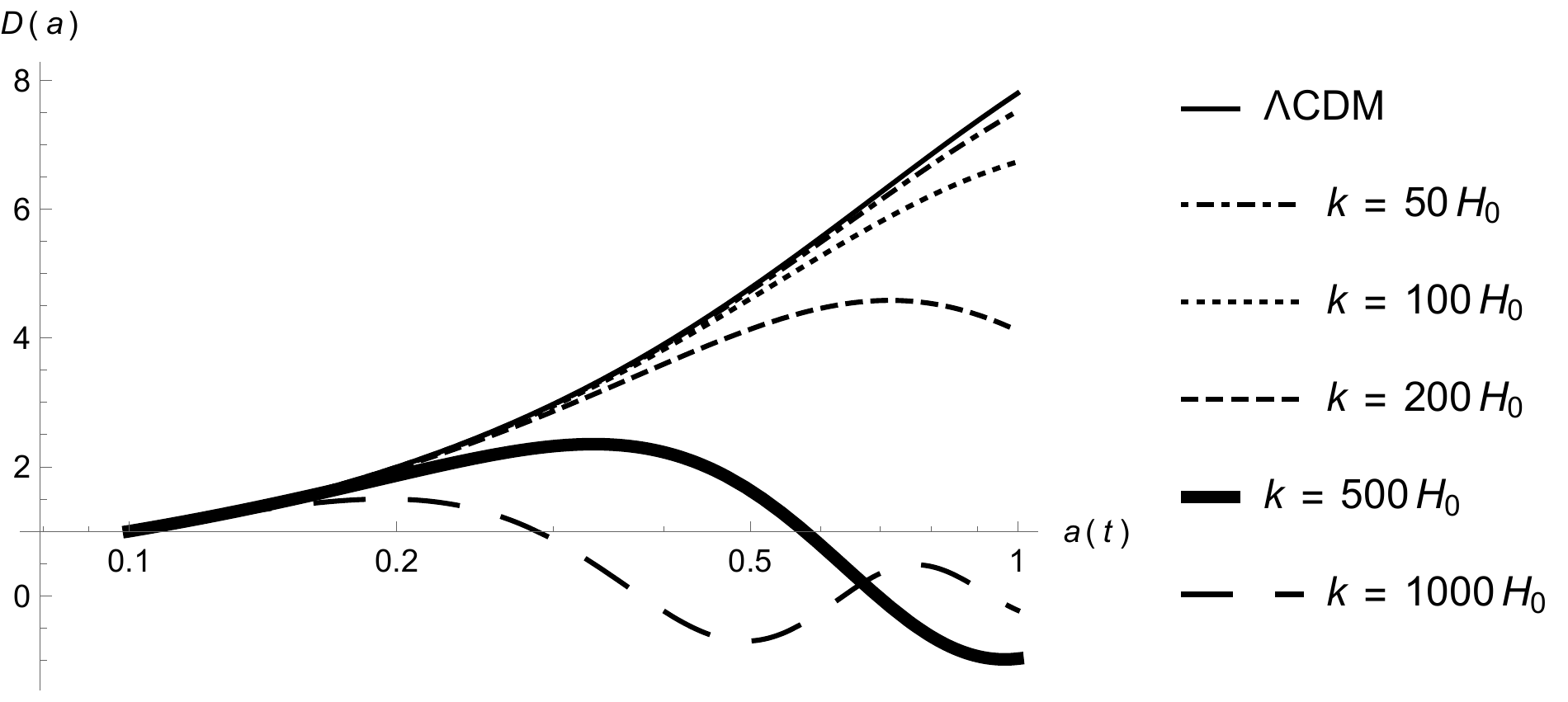}
\caption{Evolution for the $f = c_1 \sqrt{\mathcal{T}} + c_2$ model with $c_1$ and $c_2$ defined by Eqs. \eqref{eq:caseA-c1andc2-dependance} and \eqref{eq:caseA-epsilon} having $\epsilon = 10^{-4}$. Note that the solution is close to the $\Lambda$CDM solution for smaller modes but starts to deviate as $k$ is increased.}
\label{fig2}
\end{figure}

\begin{figure}[h!]
\centering
\includegraphics[width=0.49\textwidth]{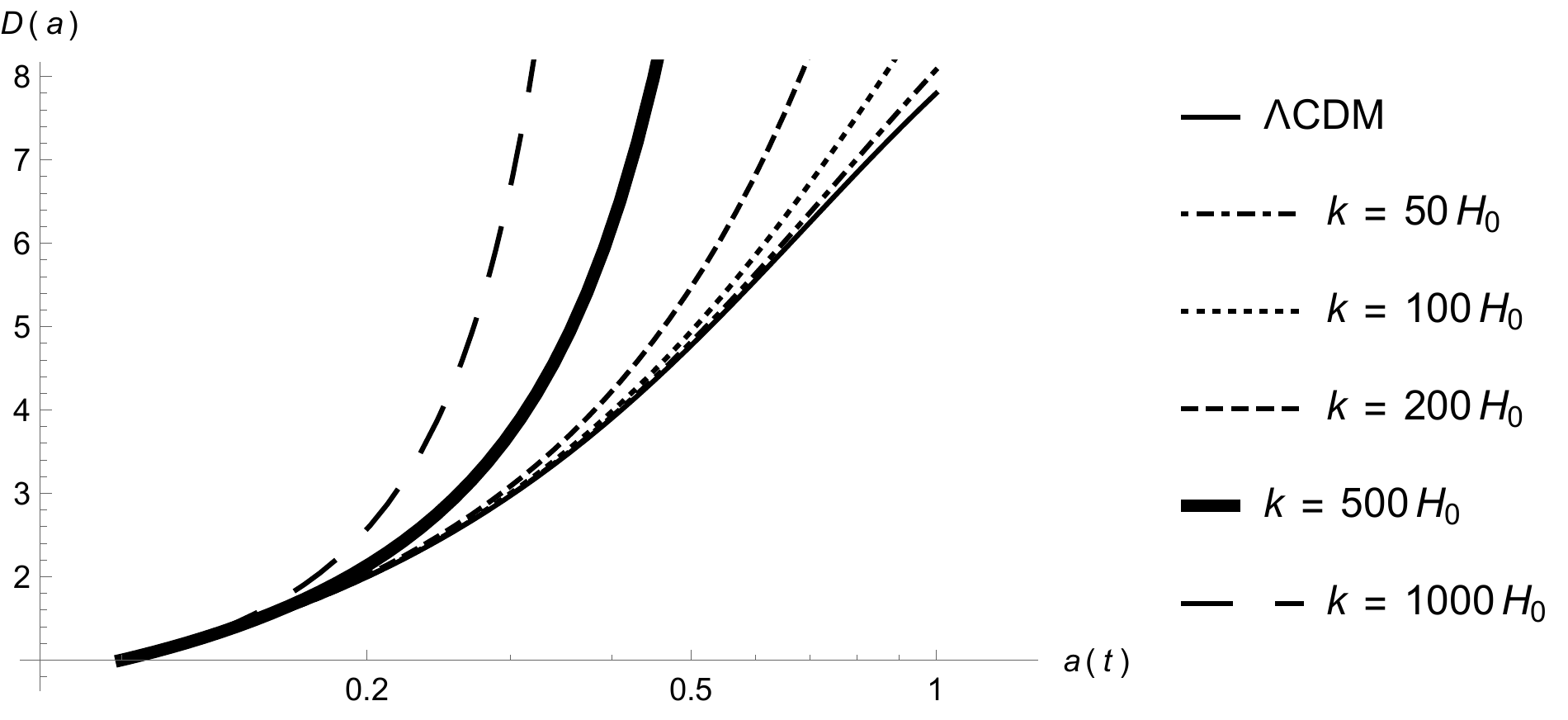}
\caption{Evolution for the $f = c_1 \sqrt{\mathcal{T}} + c_2$ model with $c_1$ and $c_2$ defined by Eqs. \eqref{eq:caseA-c1andc2-dependance} and \eqref{eq:caseA-epsilon} having $\epsilon = -10^{-4}$. Note that the solution is close to the $\Lambda$CDM solution for smaller modes but starts to deviate as $k$ is increased.}
\label{fig3}
\end{figure}

Afterwards, the non-zero cosmological constant case is considered, for two different scenarios, a positive and a negative $\epsilon$ (see Fig. \ref{fig2} and \ref{fig3}). In the former, one notes that initially, the effect of $k$ causes a slight deviation from $\Lambda$CDM ($k = 50H_0$ and $k = 100H_0$), being a slower growth rate. As the sub-horizon modes increase, this leads to much larger deviations, eventually leading to an oscillatory motion for much larger values of $k$ (compare $k = 200H_0$ with $k = 500H_0$ and $k = 1000H_0$). On the other hand, for negative $\epsilon$, a similar scenario happens for small sub-horizon modes but having a faster growth rate than $\Lambda$CDM. As $k$ increases, the growth factors increases extremely rapidly and shadows the $\Lambda$CDM evolution.

\subsubsection{II. $f = -T - \dfrac{T}{\dfrac{2c_1}{\sqrt{\mathcal{T}}}+c_2}$}

For this case, it is found that $c_1$ and $c_2$ are dependent on each other through Eq. \eqref{eq:caseB-c1andc2-dependance}, which expresses the evolution Eq. \eqref{eq:00-zero-caseB.2} by a single parameter $\epsilon$ defined in Eq. \eqref{eq:caseB-epsilon}. Hence, the evolution of $\delta_m$ can again be analysed by varying the values of $\epsilon$.

The first case considered is the extremal case $\epsilon = 1$, which gives only one type of evolution. As shown in Fig. \ref{fig6}, the effect of $k$ is already dominant, even for sufficiently small sub-horizon modes. In fact, the solution is also oscillatory, with increasing periods for larger sub-horizon modes, similar to the one found in Fig. \ref{fig1}.

\begin{figure}[h!]
\centering
\includegraphics[width=0.49\textwidth]{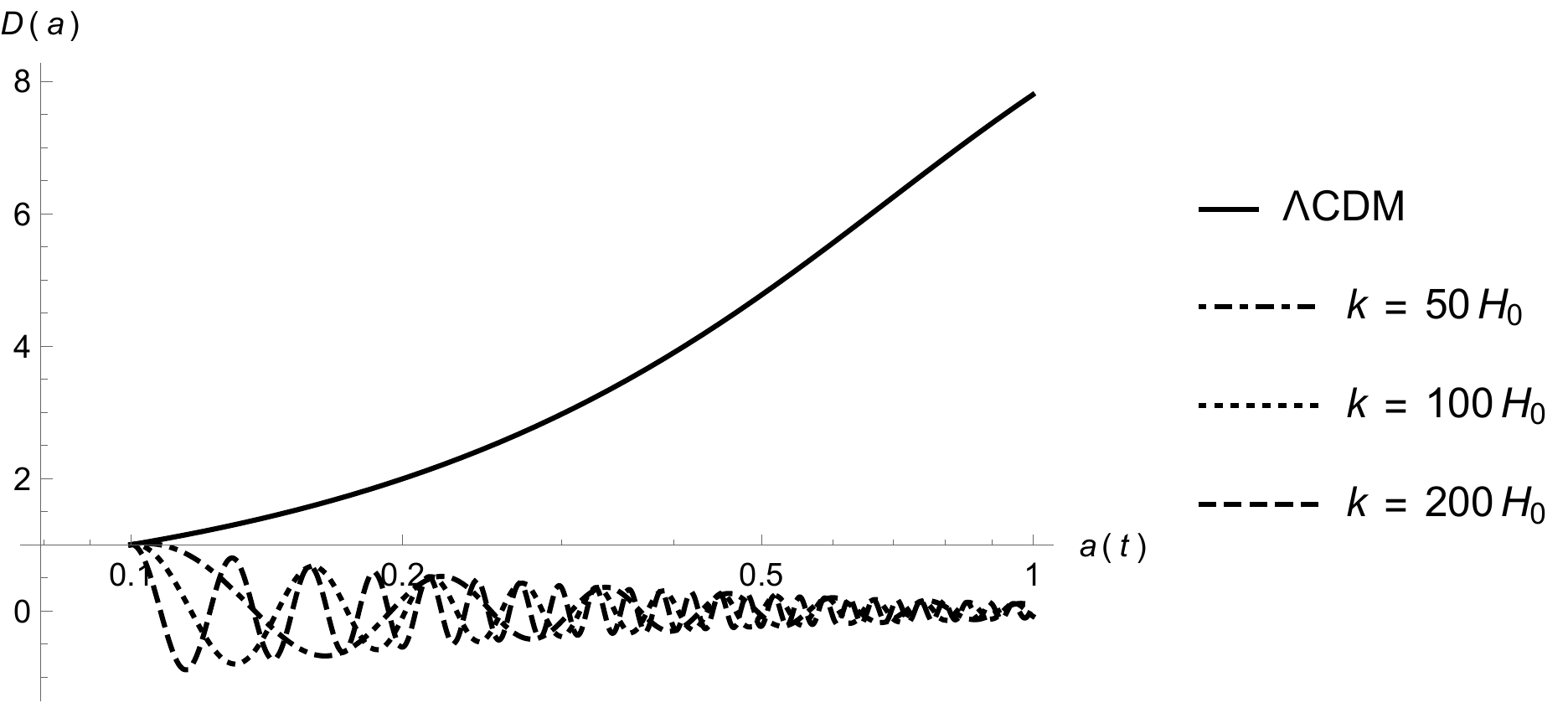}
\caption{Evolution for the $f = -T - \dfrac{T}{\dfrac{2c_1}{\sqrt{\mathcal{T}}}+c_2}$ model with $c_1$ and $c_2$ defined by Eqs. \eqref{eq:caseB-epsilon} and \eqref{eq:caseB-c1andc2-dependance} having $\epsilon = 1$. The solution deviates completely from $\Lambda$CDM and becomes oscillatory. By increasing $k$, the number of periods increase.}
\label{fig4}
\end{figure}

Subsequently, a positive and a negative $\epsilon$ was considered. In each case, this gives rise to two possible evolutions due to the presence of the plus/minus sign in Eq. \eqref{eq:00-zero-caseB.2}. Let us first consider the positive $\epsilon$ case.

\begin{figure}[h!]
\centering
\includegraphics[width=0.49\textwidth]{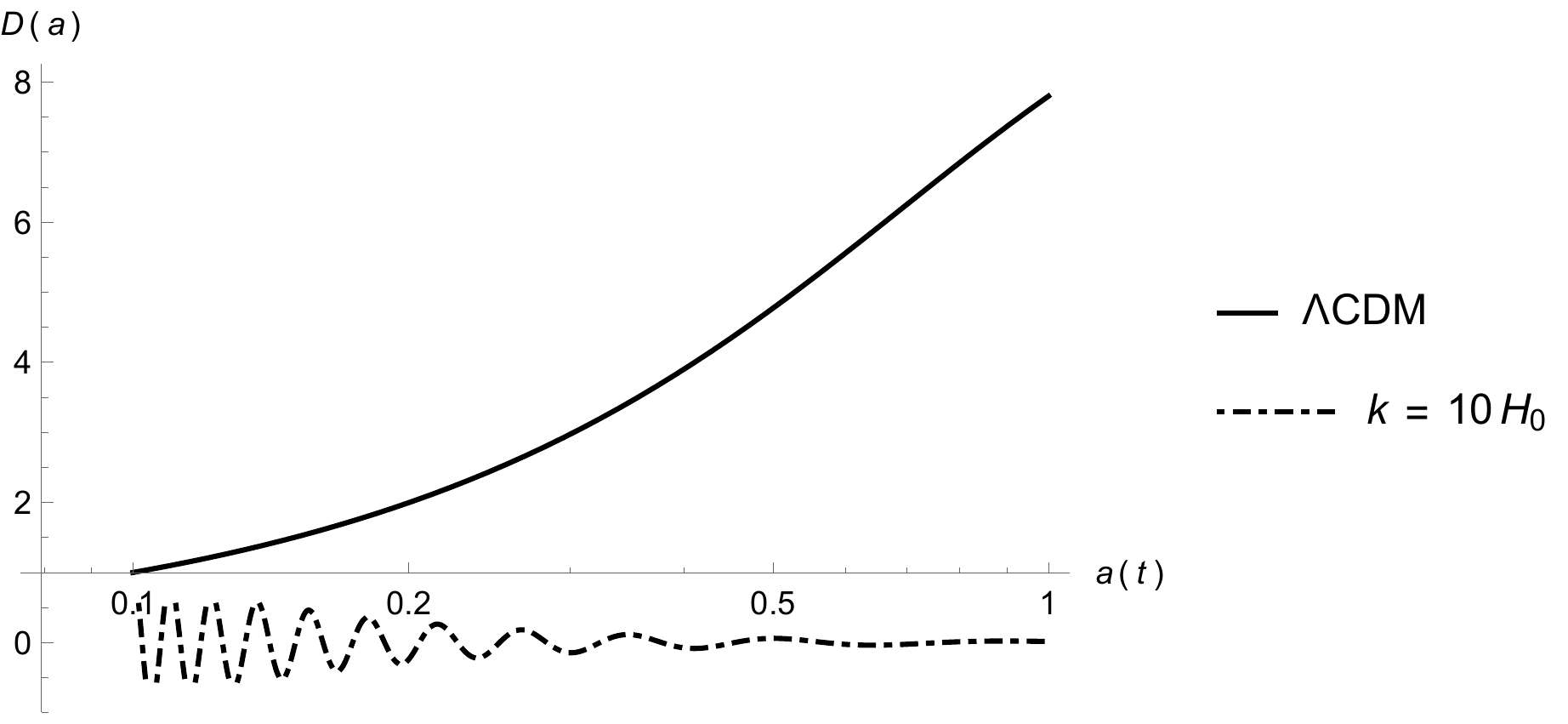}
\caption{Evolution for the $f = -T - \dfrac{T}{\dfrac{2c_1}{\sqrt{\mathcal{T}}}+c_2}$ model with $c_1$ and $c_2$ defined by Eqs. \eqref{eq:caseB-epsilon} and \eqref{eq:caseB-c1andc2-dependance} for the positive solution and $\epsilon = 10^{-3}$. The solution is oscillatory with increasing periods as $k$ increases. Note that the solution deviates greatly from $\Lambda$CDM.}
\label{fig5}
\end{figure}

\begin{figure}[h!]
\centering
\includegraphics[width=0.49\textwidth]{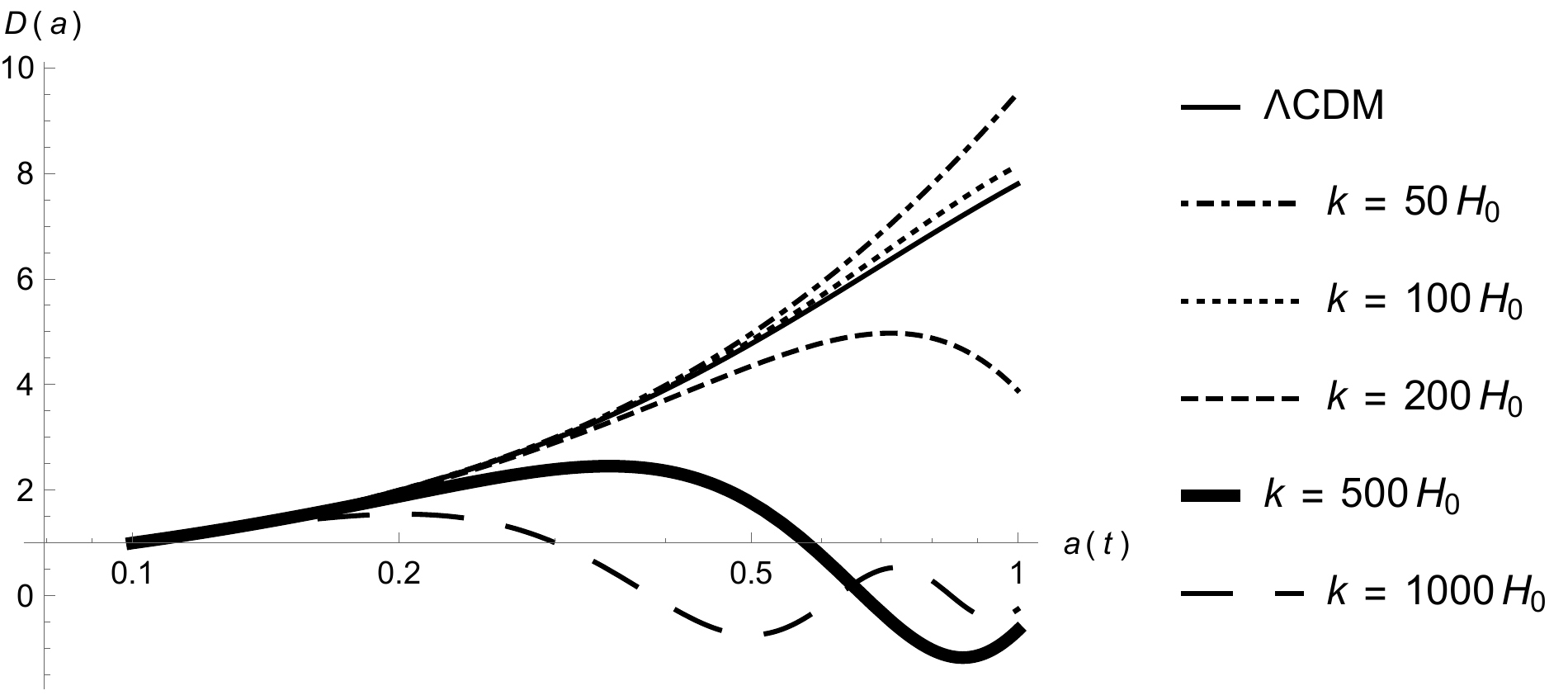}
\caption{Evolution for the $f = -T - \dfrac{T}{\dfrac{2c_1}{\sqrt{\mathcal{T}}}+c_2}$ model with $c_1$ and $c_2$ defined by Eqs. \eqref{eq:caseB-epsilon} and \eqref{eq:caseB-c1andc2-dependance} for the negative solution and $\epsilon = 10^{-3}$. The solution is close to the $\Lambda$CDM solution for smaller modes but deviates for increasing modes.}
\label{fig6}
\end{figure}

In this scenario, the effect of $k$ on the positive solution is immediately evident (Fig. \ref{fig5}). For increasing $k$, the number of periods increase extremely rapidly, making it deviate greatly from $\Lambda$CDM. On the other hand, the negative solution is somewhat close to what happens in Fig. \ref{fig2}, with the difference that the smaller modes ($k = 50H_0$ and $k = 100H_0$) have a larger growth profile than $\Lambda$CDM (Fig. \ref{fig6}).

\begin{figure}[h!]
\centering
\includegraphics[width=0.49\textwidth]{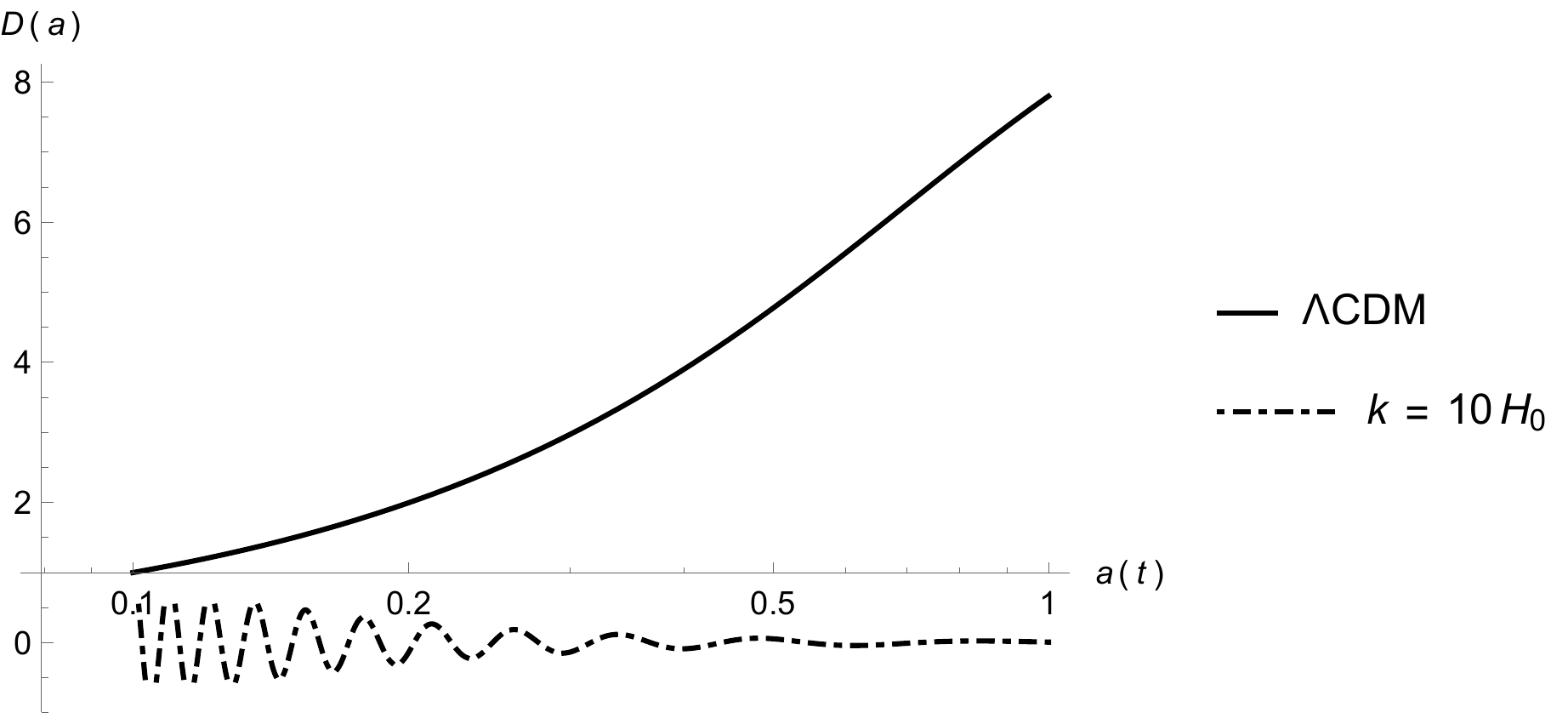}
\caption{Evolution for the $f = -T - \dfrac{T}{\dfrac{2c_1}{\sqrt{\mathcal{T}}}+c_2}$ model with $c_1$ and $c_2$ defined by Eqs. \eqref{eq:caseB-epsilon} and \eqref{eq:caseB-c1andc2-dependance} for the positive solution and $\epsilon = -10^{-3}$. The solution is oscillatory with increasing periods as $k$ increases. Note that the solution deviates greatly from $\Lambda$CDM.}
\label{fig7}
\end{figure}

\begin{figure}[h!]
\centering
\includegraphics[width=0.49\textwidth]{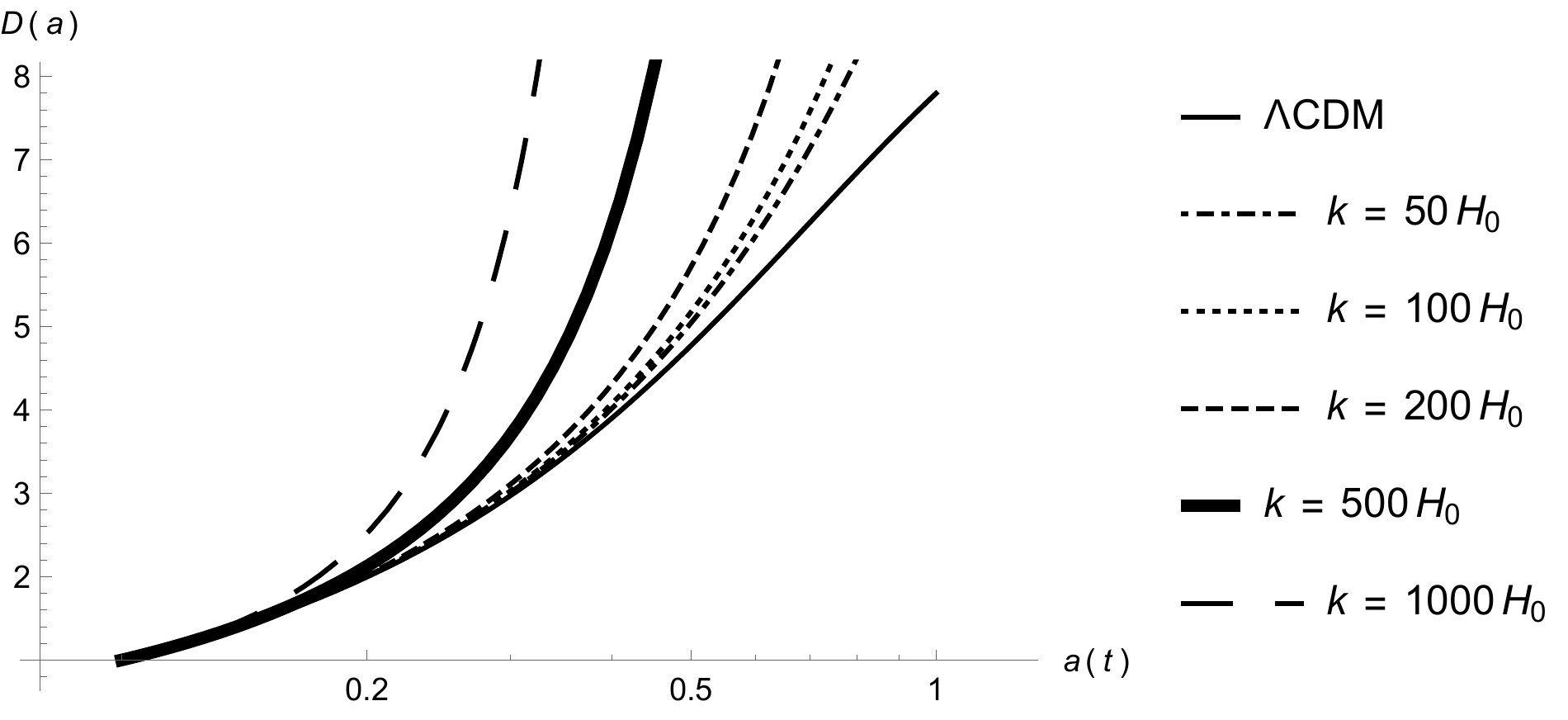}
\caption{Evolution for the $f = -T - \dfrac{T}{\dfrac{2c_1}{\sqrt{\mathcal{T}}}+c_2}$ model with $c_1$ and $c_2$ defined by Eqs. \eqref{eq:caseB-epsilon} and \eqref{eq:caseB-c1andc2-dependance} for the negative solution and $\epsilon = -10^{-3}$. The solution is close to the $\Lambda$CDM solution for smaller modes but deviates for increasing modes.}
\label{fig8}
\end{figure}

Lastly, for negative $\epsilon$, a very similar behaviour to Fig. \ref{fig5} is observed for the positive solution (Fig. \ref{fig7}). On the other hand, the negative solution is again somewhat close to what happens in Fig. \ref{fig3}, with the only difference being that the smaller modes ($k = 50H_0$ and $k = 100H_0$) have a larger growth than the ones observed in the former (Fig. \ref{fig8}). 

\subsection{B. Numerical Results for $\pi^{S} \neq 0$}

As discussed in Section III and from the results of the previous section, the $k$ dependence causes deviations from the $\Lambda$CDM solution. The only way that this can be avoided is by either neglecting both the effects of $\pi^{S}$ and $f_{\mathcal{T}}$, or having their sum be equal to zero. Since having both of them zero leads to standard $f(T)$-$\Lambda$CDM models, the second case is considered, which results in the relation Eq. \eqref{eq:pis-relation}. In this section, the case where $\pi^{S} \neq 0$ was considered, where constraints on $\pi^{S}$ were set to obtain different growth evolutions.

\subsubsection{I. $f = c_1 \sqrt{\mathcal{T}} + c_2$}

For the first model, using Eq. \eqref{eq:caseA-c1andc2-dependance}, the expression for the anisotropic term becomes
\begin{align}
\pi^{S} &= \dfrac{-\delta_m}{k^2} \dfrac{{H_0}^2 \Omega_m a^{-3}}{8\pi G} \nonumber \\
&\times \dfrac{6{H_0}^2\left(\Omega_m-1\right)-c_2}{8{H_0}^2 \Omega_m a^{-3/2}-6{H_0}^2\left(\Omega_m-1\right)+c_2}.
\end{align}
Note that by setting $\pi^{S}$ sets $c_2$, i.e. the cosmological constant (and ultimately $c_1$). Thus, we only have one free parameter, which is $\pi^{S}$. Let us consider some cases.

\paragraph{(a)} $\pi^{S} =$ constant: For this to occur, the right hand side (RHS) must become independent of $\delta_m$, $k$ and $a$. This means that $c_2$ must be dependent on these quantities. However, since $c_2$ is a constant, this becomes a contradiction. Hence, $\pi^{S}$ cannot be constant (except for 0 which reduces to $f(T)$-$\Lambda$CDM models).

\paragraph{(b)} $\pi^{S} \propto \delta_m/k^2$: Suppose that
\begin{equation}
\pi^{S} = -\gamma \dfrac{{H_0}^2 \Omega_m}{8\pi G}\dfrac{\delta_m}{k^2},
\end{equation}
where $\gamma$ is a function of time. Substituting and rearranging leads to
\begin{align}
c_2 &= \left(1+\gamma a^3\right)^{-1} \bigg\lbrace 6{H_0}^2\left(\Omega_m-1\right) \nonumber \\
&-\gamma \left[8{H_0}^2 \Omega_m a^{3/2} - 6{H_0}^2 \left(\Omega_m-1\right)a^3\right]\bigg\rbrace.
\end{align}
However, $c_2$ is a constant, and hence the RHS must become independent of time. By differentiating the expression and solving the differential equation in $\gamma$ gives a solution of the form
\begin{equation}
\gamma = \dfrac{1}{-a^3 + \eta a^{3/2}},
\end{equation}
where $\eta$ is a non-zero constant (since for $\eta = 0$, $\gamma = -a^{-3}$ leading the numerator of the RHS to still depend on time). By substituting back, we find that
\begin{equation}\label{eq:caseA-eta}
c_2 = 6{H_0}^2 \left[\Omega_m-1 -\dfrac{4\Omega_m}{3\eta}\right].
\end{equation}
One can note that this is a modification from the $\Lambda$CDM solution, provided by the last term which is only dependent on the value of $\eta$. As $|\eta| \rightarrow \infty$, the value of $c_2$ becomes the $\Lambda$CDM value [being $6{H_0}^2 \left(\Omega_m-1\right)$], which makes sense since for this case, $\gamma \rightarrow 0$ (i.e. $\pi^{S} \rightarrow 0$). Therefore, by choosing the right value for $\eta$, we can constrain the value of the cosmological constant (and hence $c_1$). Conversely, one can choose a specific form for $c_2$ (like Eq. \eqref{eq:caseA-epsilon}) to constrain the value for $\eta$ (and hence $c_1$). Since by choosing the right parameters would yield the same plots, only the constraints of $\eta$ were considered.

\begin{figure}[h!]
\centering
\includegraphics[width=0.49\textwidth]{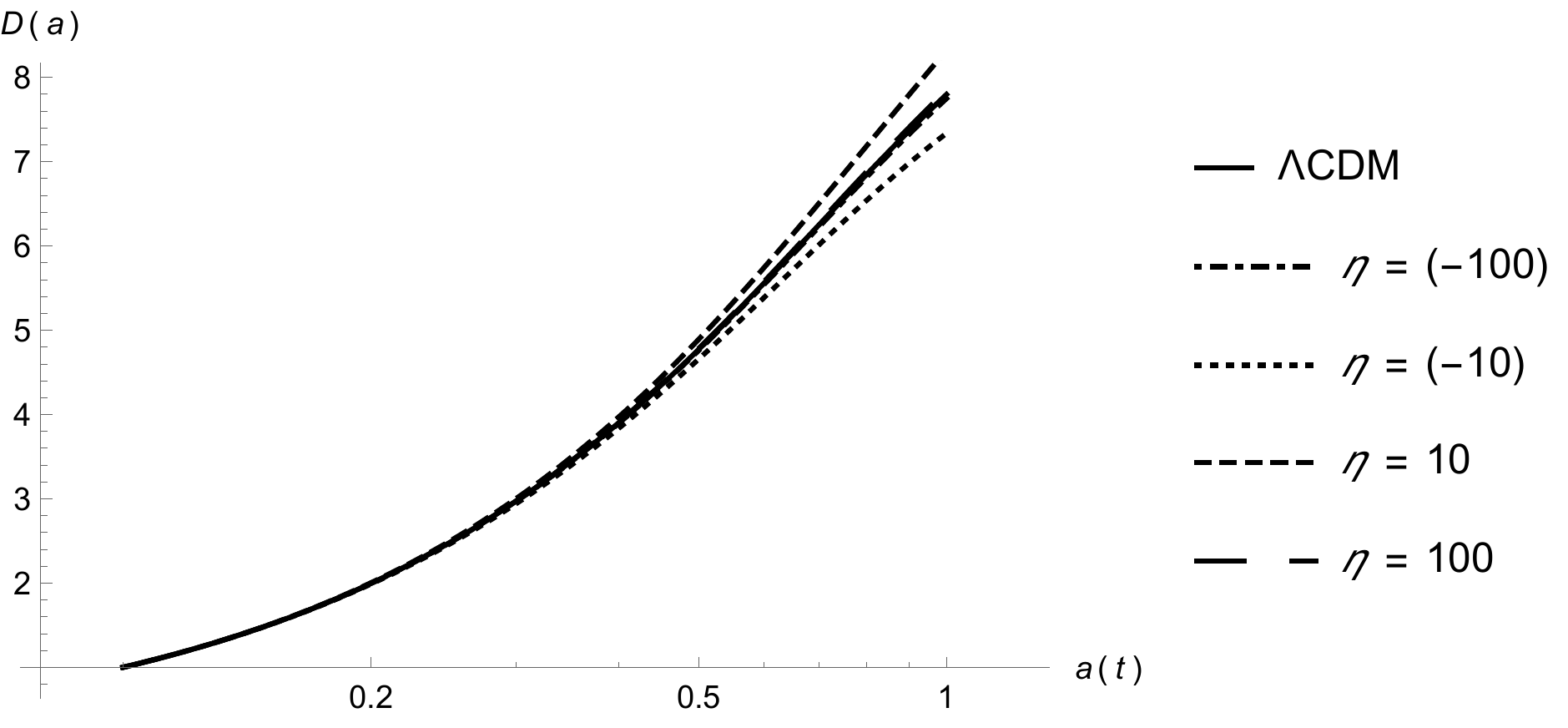}
\caption{Evolution for the $f = c_1 \sqrt{\mathcal{T}} + c_2$ model with $c_1$ and $c_2$ defined by Eqs. \eqref{eq:caseA-c1andc2-dependance} and \eqref{eq:caseA-eta} for various $\eta$. Note that the solution can be made sufficiently close to the $\Lambda$CDM solution.}
\label{fig9}
\end{figure}

In Fig. \ref{fig9}, some values of $\eta$ are considered. One can note that positive $\eta$ yields faster growths whilst negative $\eta$ yields slower growths when compared to $\Lambda$CDM. Furthermore, as the magnitude of $\eta$ increases, the closer to the $\Lambda$CDM solution gets. This is expected given the relationship between $c_2$ and $\eta$. 

\begin{figure}[h!]
\centering
\includegraphics[width=0.49\textwidth]{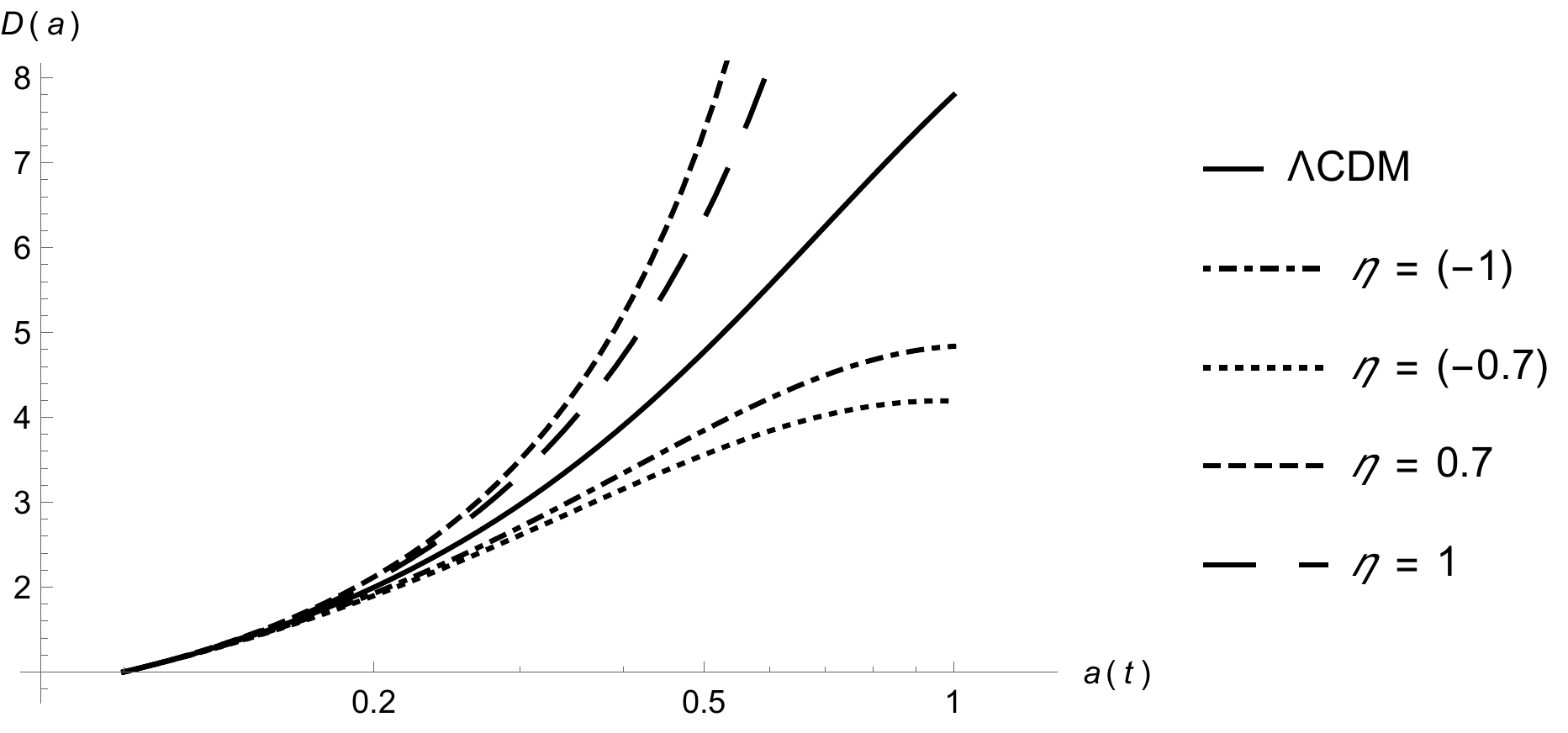}
\caption{Evolution for the $f = c_1 \sqrt{\mathcal{T}} + c_2$ model with $c_1$ and $c_2$ defined by Eqs. \eqref{eq:caseA-c1andc2-dependance} and \eqref{eq:caseA-eta} for various $\eta$. One notes that the evolutions start to deviate for values of $|\eta| < 1$. This is due to the singularity at $\eta = 0$.}
\label{fig10}
\end{figure}

On the other hand, for values of $\eta$ which are closer to zero start to deviate from the $\Lambda$CDM solution, as can be seen in Fig. \ref{fig10}. This is again due to the relationship between $c_2$ and $\eta$; for $|\eta| < 1$, the modification term starts to become large, effectively becoming large compared to the $\Lambda$CDM model value. However, the same behaviour as the previous case is retained, where positive $\eta$ yields faster growths whilst negative $\eta$ yields slower growths when compared to $\Lambda$CDM.

Thus, from the discussions about the value of $\eta$, one can easily see that $|\eta|$ has to be sufficiently large for it to be closely mimic the $\Lambda$CDM growth evolution. This imposes a constraint on the cosmological constant $c_2$, which simply states that the value should not deviate much from the $\Lambda$CDM value; otherwise the growth factor will start to either grow much faster (or much slower depending on the sign of $\eta$) than the standard $\Lambda$CDM evolution. 

\subsubsection{II. $f = -T - \dfrac{T}{\dfrac{2c_1}{\sqrt{\mathcal{T}}}+c_2}$}

In the second model, Eq. \eqref{eq:pis-relation} becomes
\begin{equation}
\pi^{S} = \dfrac{\delta_m}{k^2} \dfrac{c_1 T \mathcal{T}}{16\pi G \mathcal{T}^{1/2} \left(2c_1+c_2\sqrt{\mathcal{T}}\right)^2+3c_1 T}.
\end{equation}
Using Eq. \eqref{eq:00-zero-caseB.present-time}, the anisotropic term can be simplified into
\begin{equation}
\pi^{S} = \dfrac{\delta_m}{k^2}\dfrac{-3{H_0}^2 \mathcal{T}_0 a^{-3} c_1}{8\pi G \sqrt{\mathcal{T}_0} a^{-3/2}\left(2c_1+c_2\sqrt{\mathcal{T}_0}\right)^2 - 9 c_1 {H_0}^2}
\end{equation}
Similar to the previous case, by setting $\pi^{S}$ sets $c_1$ and $c_2$. Let us consider some cases.

\paragraph{(a)} $\pi^{S} =$ constant: For this to occur, the right hand side (RHS) must become independent of $\delta_m$, $k$ and $a$. This means that both $c_1$ and $c_2$ must be dependent on these quantities. However, since both of them are constant, this becomes a contradiction. Hence, $\pi^{S}$ cannot be constant (except for 0 which becomes the re-scaling $f(T)$ model, which is essentially a contradiction as discussed in Section IV).

\paragraph{(b)} $\pi^{S} \propto \delta_m/k^2$: Suppose that
\begin{equation}
\pi^{S} = -\gamma 6{H_0}^2 \mathcal{T}_0 \dfrac{\delta_m}{k^2},
\end{equation}
where $\gamma$ is a function of time. Substituting and rearranging leads to
\begin{align}
c_1 &= a^3 \gamma \bigg[16\pi G \sqrt{\mathcal{T}_0} a^{-3/2}\left(2c_1+c_2\sqrt{\mathcal{T}_0}\right)^2 \nonumber \\
&- 18 c_1 {H_0}^2\bigg].
\end{align}
However, $c_1$ is a constant, and hence the RHS must become independent of time. By differentiating the expression and solving the differential equation in $\gamma$ gives a solution of the form
\begin{equation}
\gamma = \dfrac{\eta a^{-3/2}}{9 a^{3/2} c_1 {H_0}^2 - 8\pi G \sqrt{\mathcal{T}_0}\left(2c_1+c_2\sqrt{\mathcal{T}_0}\right)^2},
\end{equation}
where $\eta$ is a constant. By substituting back, we find that
\begin{equation}
c_1 = -2\eta.
\end{equation}
By determining $c_1$, $c_2$ can be determined using Eq. \eqref{eq:caseB-c1andc2-dependance}, which in terms of $\eta$ gives
\begin{equation}
\eta \leq \dfrac{\sqrt{\mathcal{T}_0}}{8 \Omega_m}.
\end{equation}

This limits the possible choices of $\eta$. One can also constrain $\eta$ using $c_1$, however the plots turn out to be equivalent as long as the right constants are chosen. As was done in Section IV, one can define $\eta$ to be
\begin{equation}
\eta \equiv \epsilon\dfrac{\sqrt{\mathcal{T}_0}}{8 \Omega_m},
\end{equation}
where $\epsilon$ is a rescaling constant. Thus, the condition reduces to having $\epsilon \leq 1$. 

\begin{figure}[h!]
\centering
\includegraphics[width=0.49\textwidth]{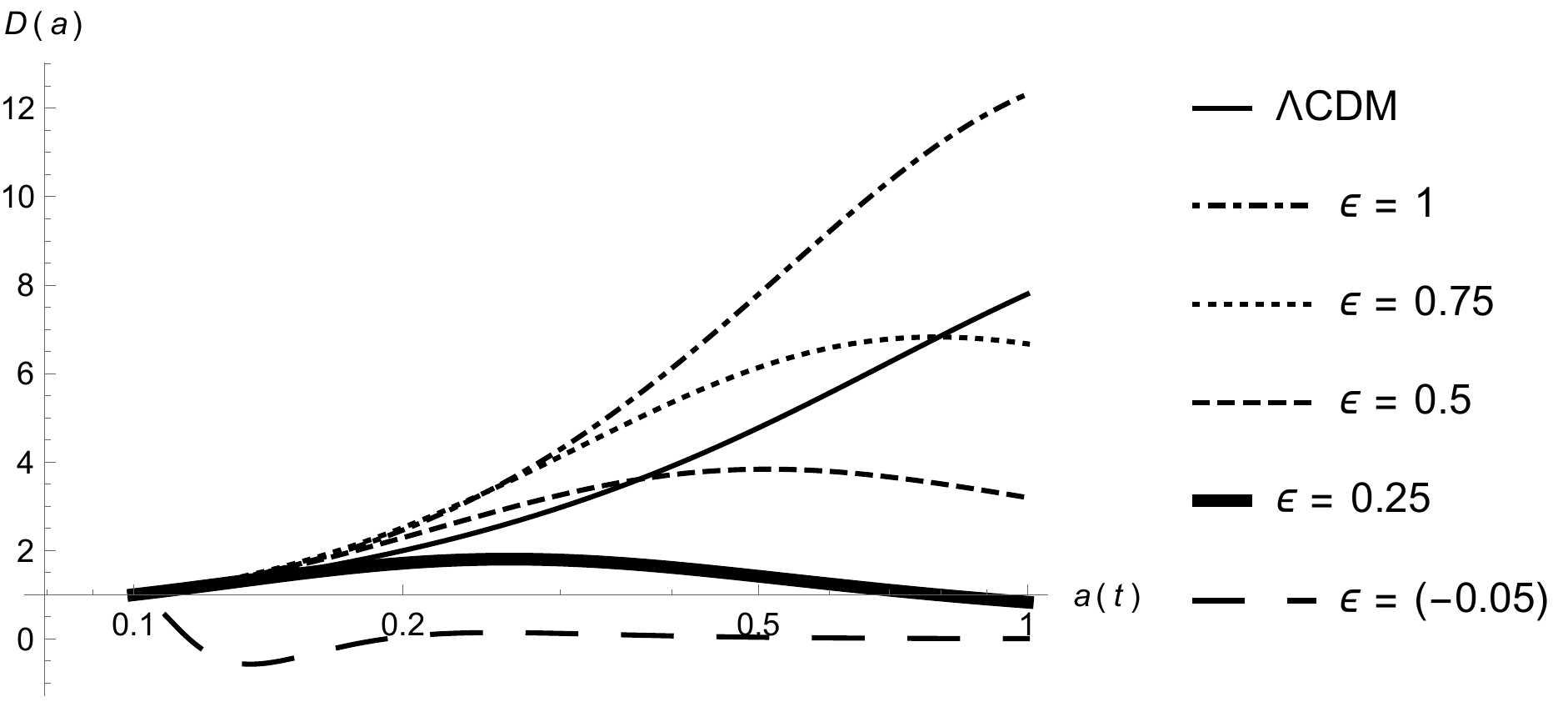}
\caption{Evolution for the $f = -T - \dfrac{T}{\dfrac{2c_1}{\sqrt{\mathcal{T}}}+c_2}$ model with $c_1$ and $c_2$ defined by Eqs. \eqref{eq:caseB-epsilon} and \eqref{eq:caseB-c1andc2-dependance} for the positive solution and various $\epsilon$. The solutions in this case deviate from the $\Lambda$CDM solution for every $\epsilon$, and becomes unstable for negative values of $\epsilon$.}
\label{fig11}
\end{figure}

In Fig. \ref{fig11}, some values of $\epsilon$ are considered. Recall that in this case, $c_2$ has two solutions and hence there can be two types of evolution. For the positive solution, one finds that no value for $\epsilon$ can describe a $\Lambda$CDM like evolution (except for $\epsilon = 1$, but with a much faster growth rate). There also seems to be a value for which the growth factors transition from being completely growing faster than $\Lambda$CDM to decreasing their growth rate and eventually fall below the latter. Furthermore, for smaller values, the solutions becomes unstable.

\begin{figure}[h!]
\centering
\includegraphics[width=0.49\textwidth]{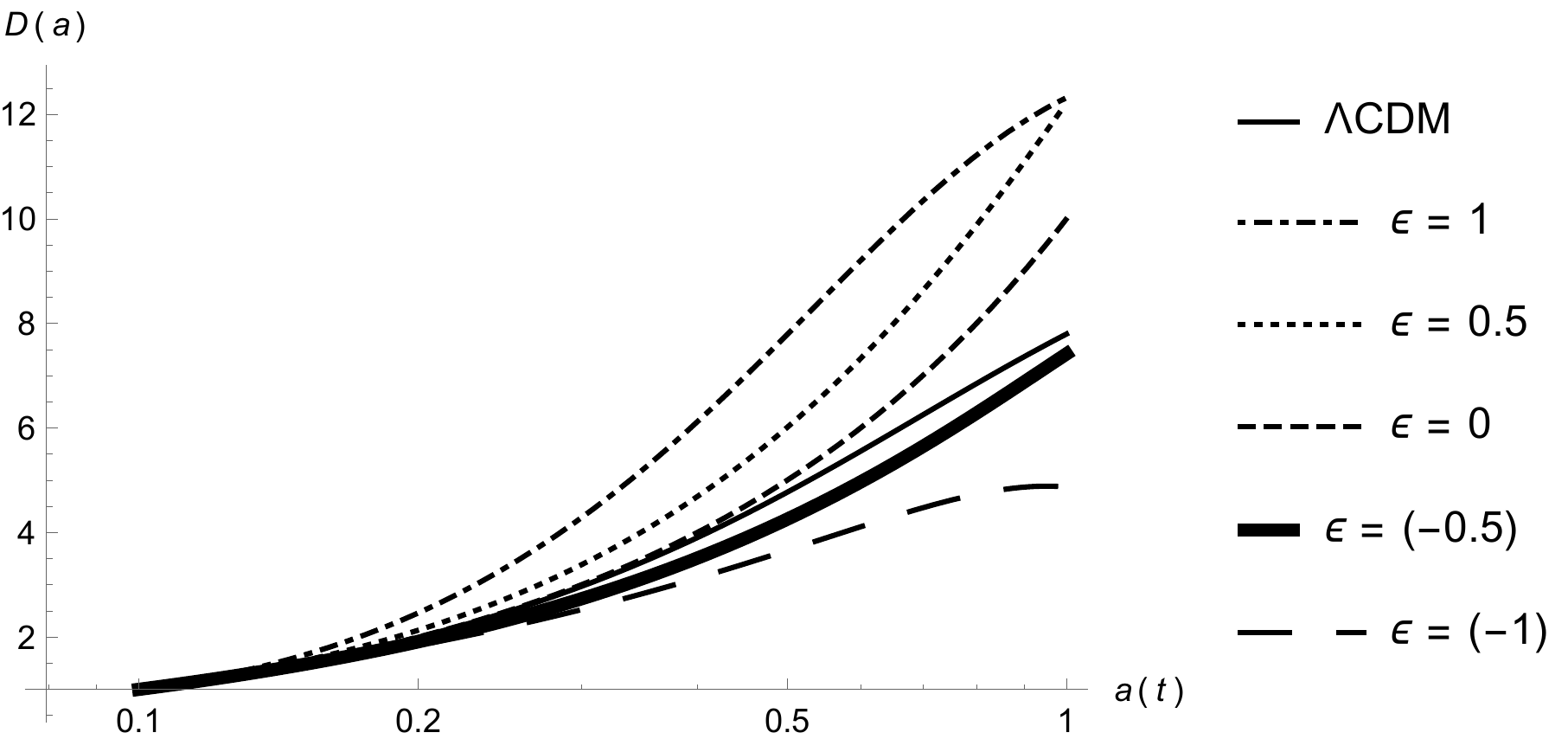}
\caption{Evolution for the $f = -T - \dfrac{T}{\dfrac{2c_1}{\sqrt{\mathcal{T}}}+c_2}$ model with $c_1$ and $c_2$ defined by Eqs. \eqref{eq:caseB-epsilon} and \eqref{eq:caseB-c1andc2-dependance} for the negative solution and various $\epsilon$. The solution can be made close to the $\Lambda$CDM solution by choosing the right $\epsilon$.}
\label{fig12}
\end{figure}

For Fig. \ref{fig12}, the negative solution for various $\epsilon$ are considered.  In this case, an evolution behaviour was noticed. For positive $\epsilon$, the growth factors are growing faster than the $\Lambda$CDM solution, while the negative values grow slower. One also notes that $\epsilon = -0.5$ evolves very similar to the $\Lambda$CDM solution, with possibly other $\epsilon$ values being closer to latter. 

\section{VI. Conclusion}\label{sec:conclusion}

The main result of the paper is the modified M\'{e}sz\'{a}ros equation for $f(T,\mathcal{T})$ gravity Eq. \eqref{eq:modified-Meszaros}. It was found that the equation depends on the sub-horizon mode $k$, which contrasts from what occurs in GR and $\Lambda$CDM models. However, this effect was also found in $f(R,T)$ models, and given the similarity between teleparallel gravity and general relativity, it is not surprising to find yet another similarity \cite{Alvarenga:2013syu}. 

In this case however, it was found that the GR limit of this modified M\'{e}sz\'{a}ros equation is still dependent on $k$ due to the presence of the anisotropic term $\pi^{S}$. Thus, this proposes two options: either this term vanishes (i.e. $\pi^{S} = 0)$ or the anisotropy is a function of the wave number such that its effect  does not cause non-physical results (for example $\pi^{S} \propto k^{n}$ with $n \leq -4$).

Even though such a scenario might exist, this would still cause a problem for $f(T,\mathcal{T})$ models due to the presence of $\dfrac{k^2 f_{\mathcal{T}}}{4a^2} \delta_m$. Since the function $f$ here is of zero order, this cannot be dependent on $k$, leading to problems. In fact, for $\pi^{S} = 0$ models, the growths found for the two functions considered provided non-physical results (either oscillatory or with varying growths which eventually deviate greatly from $\Lambda$CDM). Thus, two possible solutions were considered, the first being a constant $f$. This case would lead to standard $f(T)$-$\Lambda$CDM models (provided that the effect of $\pi^{S}$ is negligible), which is not of interest here. Thus, the second non-trivial case is considered, where the sum of this term and anisotropic term becomes zero. 

For this scenario, a coupling between the anisotropic term and the integration constants of the functions considered were found. This led to different evolutions, independent of $k$, in which some of them being close to the $\Lambda$CDM growth. Even though this might seem as a possible solution, one has to keep in mind that this sets a very specific form of how $\pi^{S}$ behaves, which can be unrealistic. The terms in the modified M\'{e}sz\'{a}ros equation have to cancel exactly, otherwise this would leave a $k$ dependence, which further strengthens this unrealistic possibility. Nonetheless, this leaves an avenue for further investigation.

This leaves $f(T,\mathcal{T})$ models being unable to explain growth evolution (except for cosmological constant models), unless the nature of the anisotropic term can be given very specific values. 

\section*{Acknowledgements}

The authors would like to thank Diego S\'{a}ez-G\'{o}mez for his comments and suggestions on an earlier version of this manuscript. The research work disclosed in this paper is partially funded by the ENDEAVOUR Scholarships Scheme.

\appendix 

\onecolumngrid

\section{Appendix I}

The action is composed of two Lagrangians to form a single Lagrangian of the form,
\begin{equation}
\mathcal{L} = \mathcal{L}_{grav} + \mathcal{L}_M,
\end{equation}
where $\mathcal{L}_{grav} \equiv e \: \left[T + f(T,\mathcal{T})\right]/16 \pi G$ and $\mathcal{L}_M \equiv e \: \mathcal{L}_m$, which denote the gravitational and matter Lagrangians respectively. The field equations are obtained by taking small variations of the action with respect to the inverse vierbein, and are found through the Euler-Lagrange equations \cite{aldrovandi2012teleparallel}
\begin{equation}
\dfrac{\partial \mathcal{L}}{\partial \udt{e}{a}{\rho}} - \partial_\sigma \dfrac{\partial \mathcal{L}}{\partial \left(\partial_\sigma \udt{e}{a}{\rho}\right)} = 0.
\end{equation}

The gravitational component of the first term is expanded as follows,
\begin{align}
16\pi G \dfrac{\partial \mathcal{L}_{grav}}{\partial \udt{e}{a}{\rho}} &= e \left[\left(1+f_T\right)\dfrac{\partial T}{\partial \udt{e}{a}{\rho}} + f_\mathcal{T} \dfrac{\partial \mathcal{T}}{\partial \udt{e}{a}{\rho}}\right] + \left[T + f(T,\mathcal{T})\right] \dfrac{\partial e}{\partial \udt{e}{a}{\rho}} \nonumber \\
&= e \left[\left(1+f_T\right)\dfrac{\partial T}{\partial \udt{e}{a}{\rho}} + f_\mathcal{T} \dfrac{\partial \left(g^{\alpha\beta} \stackrel{\textbf{em}}{T_{\alpha\beta}}\right)}{\partial \udt{e}{a}{\rho}}\right] + \left[T + f(T,\mathcal{T})\right] \dfrac{\partial e}{\partial \udt{e}{a}{\rho}} \nonumber \\
&= e \left[\left(1+f_T\right)\dfrac{\partial T}{\partial \udt{e}{a}{\rho}} + f_\mathcal{T} \left(g^{\alpha\beta}\dfrac{\partial \stackrel{\textbf{em}}{T_{\alpha\beta}}}{\partial \udt{e}{a}{\rho}}+ \stackrel{\textbf{em}}{T_{\alpha\beta}}\dfrac{\partial g^{\alpha\beta}}{\partial \udt{e}{a}{\rho}}\right)\right] + \left(T + f\right) \dfrac{\partial e}{\partial \udt{e}{a}{\rho}},
\end{align}
whilst the gravitational component of the second term is expanded to
\begin{align}
16\pi G\partial_\sigma \dfrac{\partial \mathcal{L}_{grav}}{\partial \left(\partial_\sigma \udt{e}{a}{\rho}\right)} &= \partial_\sigma \left[e \left(1+f_T\right) \dfrac{\partial T}{\partial \left(\partial_\sigma \udt{e}{a}{\rho}\right)} + e f_\mathcal{T} \dfrac{\partial \mathcal{T}}{\partial \left(\partial_\sigma \udt{e}{a}{\rho}\right)} \right] \nonumber \\
&= \left(1+f_T\right) \partial_\sigma \left[e \dfrac{\partial T}{\partial \left(\partial_\sigma \udt{e}{a}{\rho}\right)}\right] + e \dfrac{\partial T}{\partial \left(\partial_\sigma \udt{e}{a}{\rho}\right)} f_{TT} \partial_\sigma T +  f_\mathcal{T} \partial_\sigma \left[e \dfrac{\partial \mathcal{T}}{\partial \left(\partial_\sigma \udt{e}{a}{\rho}\right)}\right] \nonumber \\
&+ e \dfrac{\partial \mathcal{T}}{\partial \left(\partial_\sigma \udt{e}{a}{\rho}\right)} f_{\mathcal{T}\mathcal{T}} \partial_\sigma \mathcal{T} \nonumber \\
&= \left(1+f_T\right) \partial_\sigma \left[e \dfrac{\partial T}{\partial \left(\partial_\sigma \udt{e}{a}{\rho}\right)}\right] + e \dfrac{\partial T}{\partial \left(\partial_\sigma \udt{e}{a}{\rho}\right)} f_{TT} \partial_\sigma T +  f_\mathcal{T} \partial_\sigma \left[e \dfrac{\partial \mathcal{T}}{\partial \left(\partial_\sigma \udt{e}{a}{\rho}\right)}\right] \nonumber \\
&+ e \dfrac{\partial T}{\partial \left(\partial_\sigma \udt{e}{a}{\rho}\right)} f_{T\mathcal{T}} \partial_\sigma \mathcal{T}.
\end{align}

On the other hand, under the assumption that the matter Lagrangian does not depend on the derivatives of the inverse vierbein field, the matter components of the Euler-Lagrange terms become,
\begin{align}
\dfrac{\partial \mathcal{L}_{M}}{\partial \udt{e}{a}{\rho}} &= \dfrac{\partial \left(e \mathcal{L}_m\right)}{\partial \udt{e}{a}{\rho}}, \\
\partial_\sigma \dfrac{\partial \mathcal{L}_{M}}{\partial \left(\partial_\sigma \udt{e}{a}{\rho}\right)} &= 0.
\end{align}
This condition simplifies the second gravitational term to 
\begin{align}
16\pi G\partial_\sigma \dfrac{\partial \mathcal{L}_{grav}}{\partial \left(\partial_\sigma \udt{e}{a}{\rho}\right)} &= \left(1+f_T\right) \partial_\sigma \left[e \dfrac{\partial T}{\partial \left(\partial_\sigma \udt{e}{a}{\rho}\right)}\right] + e \dfrac{\partial T}{\partial \left(\partial_\sigma \udt{e}{a}{\rho}\right)} f_{TT} \partial_\sigma T + e \dfrac{\partial T}{\partial \left(\partial_\sigma \udt{e}{a}{\rho}\right)} f_{T\mathcal{T}} \partial_\sigma \mathcal{T}.
\end{align}
As given in Refs. \cite{aldrovandi2012teleparallel,Krssak:2015oua}, the following relations for the derivatives are given,
\begin{align}
\dfrac{\partial T}{\partial \left(\partial_\sigma \udt{e}{a}{\rho}\right)} &= -4 \dut{S}{a}{\rho\sigma}, \\
\dfrac{\partial T}{\partial \udt{e}{a}{\rho}} &= -4\udt{T}{b}{\nu a}\dut{S}{b}{\nu\rho} + 4 \udt{\omega}{b}{a\nu}\udt{S}{b}{\nu\rho}, \\
\dfrac{\partial e}{\partial \udt{e}{a}{\rho}} &= e \dut{e}{a}{\rho}, \\
\dfrac{\partial g^{\alpha\beta}}{\partial \udt{e}{a}{\rho}} &= -g^{\rho\beta}\dut{e}{a}{\alpha}-g^{\rho\alpha}\dut{e}{a}{\beta}.
\end{align}
Therefore, the field equations for $f\left(T,\mathcal{T}\right)$ gravity become
\begin{align}
&\left(1+f_T\right)\left[e^{-1} \partial_\sigma \left(e \dut{S}{a}{\rho\sigma}\right)-\udt{T}{b}{\nu a}\dut{S}{b}{\nu\rho} + \udt{\omega}{b}{a\nu}\udt{S}{b}{\nu\rho}\right]
+ \left(f_{TT} \partial_\sigma T + f_{T\mathcal{T}} \partial_\sigma \mathcal{T}\right)\dut{S}{a}{\rho\sigma}
+ \dut{e}{a}{\rho} \left(\dfrac{T + f}{4}\right) \nonumber \\
&+ \dfrac{f_\mathcal{T}}{4} \left[g^{\alpha\beta}\dfrac{\partial \stackrel{\textbf{em}}{T_{\alpha\beta}}}{\partial \udt{e}{a}{\rho}}+ \stackrel{\textbf{em}}{T_{\alpha\beta}}\left(-g^{\rho\beta}\dut{e}{a}{\alpha}-g^{\rho\alpha}\dut{e}{a}{\beta}\right)\right]
= -4\pi G e^{-1} \dfrac{\partial \left(e \mathcal{L}_m\right)}{\partial \udt{e}{a}{\rho}}.
\end{align}
By defining the stress-energy tensor to be,
\begin{equation}
\stackrel{\textbf{em}}{\dut{T}{a}{\rho}} \equiv - e^{-1} \dfrac{\partial \left(e \mathcal{L}_m\right)}{\partial \udt{e}{a}{\rho}},
\end{equation}
and considering a perfect fluid representation, we have \cite{Harko:2011kv}
\begin{equation}
g^{\alpha\beta}\dfrac{\partial \stackrel{\textbf{em}}{T_{\alpha\beta}}}{\partial \udt{e}{a}{\rho}} = 4\stackrel{\textbf{em}}{\dut{T}{a}{\rho}} + 2p\dut{e}{a}{\rho}.
\end{equation}
Therefore, the final field equations are given to be,
\begin{align}
&\left(1+f_T\right)\left[e^{-1} \partial_\sigma \left(e \dut{S}{a}{\rho\sigma}\right)-\udt{T}{b}{\nu a}\dut{S}{b}{\nu\rho} + \udt{\omega}{b}{a\nu}\udt{S}{b}{\nu\rho}\right]
+ \left(f_{TT} \partial_\sigma T + f_{T\mathcal{T}} \partial_\sigma \mathcal{T}\right)\dut{S}{a}{\rho\sigma}
+ \dut{e}{a}{\rho} \left(\dfrac{T + f}{4}\right) \nonumber \\
&+ \dfrac{f_\mathcal{T}}{2} \left(\stackrel{\textbf{em}}{\dut{T}{a}{\rho}} + p\dut{e}{a}{\rho}\right)
= 4\pi G \stackrel{\textbf{em}}{\dut{T}{a}{\rho}}.
\end{align}

\section{Appendix II}

\noindent The components of the vierbein $\dut{e}{\mu}{A}$ are given to be,
\begin{align*}
&\dut{e}{0}{0} = 1 + \phi, && \dut{e}{i}{0} = a \partial_i \tilde{w}, \\
&\dut{e}{0}{i} = -\partial^i w, && \dut{e}{j}{i} = a \left[(1-\psi)\delta^{i}_{j} - \partial_j \partial^i h - \dut{\epsilon}{j}{in}\partial_n \tilde{h}\right],
\end{align*}
while the inverse $\udt{e}{\mu}{A}$ by,
\begin{align*}
&\udt{e}{0}{0} = 1 - \phi, && \udt{e}{i}{0} = a^{-1} \partial^i w, \\
&\udt{e}{0}{i} = -\partial_i \tilde{w}, && \udt{e}{i}{j} = a^{-1} \left[(1+\psi)\delta^{i}_{j} + \partial_j \partial^i h + \dut{\epsilon}{j}{in}\partial_n \tilde{h}\right].
\end{align*}
To obtain the perturbed FLRW metric, a Newtonian gauge is considered, which sets $w = -\tilde{w}$, $h = 0$. Under this gauge, the non-zero components of the torsion tensor components are given by
\begin{align*}
\udt{T}{0}{0i} &= -\partial_i \phi - a \partial_i \dot{w}, \\
\udt{T}{0}{ij} &= 0, \\
\udt{T}{i}{0j} &= H \delta^i_j - \dot{\psi}\delta^i_j - \dut{\epsilon}{j}{in}\partial_n \dot{\tilde{h}} + a^{-1} \partial_j \partial^i w, \\
\udt{T}{i}{jk} &= \partial_k \psi \delta^i_j - \partial_j \psi \delta^i_k + \dut{\epsilon}{j}{in}\partial_k \partial_n \tilde{h} - \dut{\epsilon}{k}{in}\partial_j \partial_n \tilde{h},
\end{align*}
\noindent while the non-zero superpotential tensor components are
\begin{align*}
\dut{S}{0}{0i} &= a^{-2} \partial^i \psi, \\
\dut{S}{0}{ij} &= -\dfrac{1}{2}a^{-2} \epsilon^{ijn}\partial_n \dot{\tilde{h}}, \\
\dut{S}{i}{0j} &= -H\delta^j_i + \left(\dot{\psi} + 2H\phi - \dfrac{1}{2} a^{-1} \partial^2 w\right) \delta^j_i + \dfrac{1}{2} a^{-1} \partial_i \partial^j w, \\
\dut{S}{i}{jk} &= -\dfrac{1}{2} a^{-2} \epsilon^{jkn}\partial_i \partial_n \tilde{h} + \dfrac{1}{2}a^{-2}\left(\partial^k \psi - \partial^k \phi - a \partial^k \dot{w} \right) \delta^j_i - \dfrac{1}{2}a^{-2} \left(\partial^j \psi - \partial^j \phi - a\partial^j \dot{w}\right) \delta^k_i.
\end{align*}
Using Eq. \eqref{eq:torsionscalardef},  the torsion scalar is found to be,
\begin{equation*}
T = -6H^2 + 12H\left(\dot{\psi} + H\phi\right) - 4a^{-1}H\partial^2 w.
\end{equation*}
Due to the signature used in this paper, the stress energy tensor $\stackrel{\textbf{em}}{\dut{T}{\alpha}{\rho}}$ takes the form,
\begin{equation*}
\stackrel{\textbf{em}}{T^{\mu\nu}} = (\rho + p) u^\mu u^\nu - pg^{\mu\nu} - \Pi^{\mu\nu},
\end{equation*}
where $u^\mu$ is the fluid four velocity, and $\Pi^{\mu\nu}$ is the anisotropic stress tensor which satisfies the following properties, $\udt{\Pi}{0}{0} = \udt{\Pi}{0}{i} = u^{\mu} \Pi_{\mu\nu} = 0$. For the veirbein considered, the components of the stress-energy tensor are given by,
\begin{align*}
&\stackrel{\textbf{em}}{\dut{T}{0}{0}} = \rho + \delta\rho, && \stackrel{\textbf{em}}{\dut{T}{0}{i}} = \left(\rho + p\right) \partial^i v, \\
&\stackrel{\textbf{em}}{\dut{T}{i}{0}} = -a^2 \left(\rho + p\right)\partial_i v, && \stackrel{\textbf{em}}{\dut{T}{i}{j}} = -\left(p + \delta p\right) \delta^j_i - \partial_i \partial^j \pi^{\text{S}}.
\end{align*}
where $v$ and $\pi^{S}$ are the scalar components of the velocity vector $v^{i} \equiv u^i/u^0$ and the anisotropic stress. Hence, the trace $\mathcal{T}$ is
\begin{equation*}
\mathcal{T} = \rho + \delta\rho -3 \left(p + \delta p\right) - \partial^2 \pi^{\text{S}}.
\end{equation*}

\twocolumngrid

\end{document}